\documentclass{aa}
\usepackage{psfig}
\usepackage{latexsym}

\newcommand{\cts}{cts\,s$^{-1}$}

\newcommand{\msun}{M_{\odot}}

\newcommand{\ltap}{\mathrel{\hbox{\rlap{\lower.55ex \hbox {$\sim$}}
                   \kern-.3em \raise.4ex \hbox{$<$}}}}
\newcommand{\gtap}{\mathrel{\hbox{\rlap{\lower.55ex \hbox {$\sim$}}
                   \kern-.3em \raise.4ex \hbox{$>$}}}}
\newcommand{\nh}{N_{\rm H}}
\newcommand{\lx}{L_{\rm x}}
\newcommand{\cmsq}{cm$^{-2}$}
\newcommand{\ctks}{cts\,ksec$^{-1}$}
\newcommand{\ergs}{erg\,s$^{-1}$}
\newcommand{\ergcms}{erg\,cm$^{-2}$s$^{-1}$}

\begin{document}
\thesaurus{05(10.07.3, 13.25.5)}
\title{A census with ROSAT of low-luminosity X-ray sources in globular clusters
}
\author{F.~ Verbunt }
\offprints{F.~Verbunt}
\mail{verbunt@phys.uu.nl}

\institute{     Astronomical Institute,
              P.O.Box 80000, NL-3508 TA Utrecht, The Netherlands
                        }
\date{\today }   
\maketitle


\begin{abstract}
I analyze 101 observations from the ROSAT archive to search for X-ray sources 
in or near 55 globular clusters. New sources are found in the cores of 
NGC\,362 (a double source), NGC\,6121 (marginally significant), NGC 6139, and
NGC\,6266; and outside the cores of NGC\,6205, NGC\,6352 and NGC\,6388.
More accurate positions are determined for the X-ray sources in some
ten clusters.
The improved position for the source in  NGC\,6341 excludes the suggested
ultraviolet counterpart. It is shown that one of the two sources reported
near the core NGC\,6626 is spurious, as is the detection of a pulsar
period in the PSPC data of this cluster; the central source is
resolved in three sources.
One source reported previously in NGC\,6304 is demoted to an upper limit.
For 20 cluster cores better upper limits to the X-ray luminosity are
obtained.

From a statistical analysis I argue that several sources outside the cluster
cores may well belong to the clusters. All spectral energy distributions
observed so far are relatively soft, with bremsstrahlung temperatures
$\simeq0.9$\,keV; there is evidence however that bremsstrahlung spectra do not
correctly describe the spectra. The X-ray luminosity per unit mass for the 
cluster as a whole does not depend on the concentration; the luminosity
per unit mass for the core may increase with the cluster concentration.

\end{abstract}

\section{Introduction}

There are a few hundred bright X-ray sources in the whole sky, with
countrates in excess of 0.1\,\cts in early satellites as UHURU, and
in excess of 10\,\cts in ROSAT; there are dozens of sources per
square degree at the lowest countrates (about 0.5 cts\,ks$^{-1}$)
detectable by ROSAT in a long observation.  Whenever an X-ray source
is detected in the direction of an extended object like a globular
cluster, the question must be faced whether the source actually
belongs to the cluster or is a fore- or background object.  The
answer to the question depends on the flux of the source and on the
extent of the search area.  Thus, a bright source projected against the
small core of a globular cluster may be safely asserted to be
associated with the cluster, but a faint source projected against the
outskirts of the cluster is more likely to be a foreground star or
background active galaxy.  At very low flux levels even a source in
the cluster core may be a chance projection.

Whereas the twelve bright sources detected in the direction of as many
globular clusters are clearly members of these clusters (for a list
see e.g. Hut et al.\ 1992), the situation is more complicated for the
fainter sources.  For example, Hertz \&\ Grindlay (1983) detected five
sources in the direction of $\omega$\,Cen, one projected on the core,
and four outside it. Three of the sources detected outside the core
have been identified with foreground stars, certainly not cluster
members (Margon \&\ Bolte 1987, Cool et al.\ 1995); the fourth one may
well be a fore- or background object as well (Verbunt et al.\ 1995).
With ROSAT three sources are detected in projection against the large
core of $ \omega$\,Cen, the brighter two probably belonging to the
cluster, but the faintest one possibly a fore- or background object
(Verbunt \&\ Johnston 2000).  It would be rash, however, to conclude
that all faint sources outside the cluster core are not related to
the cluster. Rappaport et al.\ (1994) detect three sources in the
$40'\times40'$ field of view of a ROSAT HRI observation of the
globular cluster Pal\,2, one of which has a distance of 32$''$ to the cluster
centre; even though this source is well outside the core radius of
this cluster, and even taking into account the higher sensitivity near
the center of the field of view, the probability that a single
serendipitous source is this close to the cluster is small, and
Rappaport et al.\ argue that the source is a cluster member. 
The dwarf nova V\,101 is well outside the core radius, and even outside 
the half-mass radius, of the globular cluster NGC\,5904, but is detected 
in X-rays and most probably is a member of the cluster (Hakala et al.\ 1997).
\nocite{hmg+92}\nocite{hg83}\nocite{mb87}\nocite{cgb+95}\nocite{vbhj95}
\nocite{vj00}\nocite{rdlm94}\nocite{hcjv97}

A census of X-ray sources outside the cores of but belonging to
globular clusters is interesting for two main reasons.  Such sources
may belong to the population of primordial binaries in globular
clusters that may show up as an X-ray source. The cataclysmic variable
V\,101 in NGC\,5904 may well be an example. The population of X-ray
sources in the cores of globular clusters must be studied in
comparison with this primordial population: in a relatively open
globular cluster most binaries in the core are primordial
(see e.g.\ Verbunt \&\ Meylan 1988, Davies 1997).
On the other hand, some of the sources outside the cores may originate
in the cores, being catapulted out of them in three-body
interactions; at the moment no examples of such X-ray sources can be
identified, but that such ejection does occur is shown by the radio
pulsar binary C in M\,15 (Sigurdsson \&\ Phinney 1993). These ejected
sources probably are a minor fraction, but they are interesting as
providing information of the nature and frequency of close encounters
between binaries and single stars or between binaries and binaries in
the cluster cores, processes that are instrumental in the evolution of
the cluster.

The first goal of this paper is to provide a census of low-luminosity
X-ray sources in globular clusters, both in their cores and outside
them, on the basis of the observations made with ROSAT.  For many
sources only optical identification will decide whether they are
members of the cluster or not, and a secondary goal of this paper is
to obtain accurate positions for the X-ray sources where possible.  To
obtain a homogeneous data base -- for example with respect to the
assumed size of the ROSAT HRI pixels, see Sect.\,2 -- , I do not only
analyze previously unpublished data, but also re-analyse some
previously published data.  Because low-luminosity sources cannot be
detected in clusters with bright X-ray sources, this paper is limited
to those clusters in which no bright source is detected, but includes 
clusters with X-ray transients in their low state. The data analysis and
selection is described in Sect.\,2, the results are give in Sect\,3,
and a first analysis and discussion of the results is given in Sect.\,4.

\section{Data analysis}

\begin{table*}[]
\caption[o]{ROSAT observations of globular clusters. For each observation
I indicate which detector was used, H(RI) or P(SPC), the month, beginning and
end of the observation, the effective exposure time, and where applicable
an offset. There are three types of offset indicated, as follows:
a single number of arcminutes indicates the distance of the cluster center
to the pointing direction (i.e.\ the globular cluster is serendipitously
in the field of view of an observation of another target), two integers
indicate the shift in $x,y$ pixels applied before adding the observation to
other observations of the same cluster (one pixel is almost $0.5''$,
note that a positive pixel shift corresponds to lower right ascension and
lower declination), two numbers in (arc)seconds indicate the correction 
applied to bring the ROSAT X-ray coordinate frame into agreement with the 
optical J\,2000 system (only possible when at least one X-ray source has an 
optical counterpart).
\label{tobse}}
\begin{tabular}
{l@{\hspace{0.2cm}}l@{\hspace{0.2cm}}l@{\hspace{0.2cm}}r@{\hspace{0.2cm}}rll@{\hspace{0.2cm}}l@{\hspace{0.2cm}}l@{\hspace{0.2cm}}r@{\hspace{0.2cm}}rl}
no. & \multicolumn{3}{c}{\mbox observed (JD$-$2440000)} & $t_{\rm exp}$(s) &
offset &
no. & \multicolumn{3}{c}{\mbox observed (JD$-$2440000)} & $t_{\rm exp}$(s) &
offset \\
   104 & P & 91/03 &  8343.741--352.676 & 20148 &  &                              
  6341 & P & 92/08 &  8861.995--865.548 & 45541 & \\                              
   104 & P & 92/04 &  8730.616--751.255 & 34028 &  &                              
  6352 & P & 93/09 &  9254.788--256.390 &  9465 & $-0\fs37,0\farcs$0\\            
   362 & H & 95/05 &  9851.086--854.962 & 20060 &  &                              
  6366 & P & 93/08 &  9229.414--229.965 &  9807 & $+0\fs18,+1\farcs0$\\           
  1261 & P & 93/01 &  8983.709--024.779 & 45766 & $25'$  &                        
 HP\,1 & H & 93/03 &  9065.584--065.604 &  1747 & \\                              
  1261 & H & 95/02 &  9750.835--751.850 &  5478 & $+7$,$+4$  &                    
  6380 & H & 92/03 &  8696.302--698.182 &  4858 & \\                              
  1261 & H & 95/11 & 10049.573--050.576 &  3145 &  &                              
  6388 & P & 91/09 &  8526.894--527.832 & 20937 & 33$'$ ($-0\fs34,+0\farcs9$) \\  
Pal\,2 & H & 92/03 &  8694.469--697.019 &  6194 &  &                              
  6388 & P & 92/09 &  8894.048--894.065 &  1510 & 26$'$ ($+0\fs55,-0\farcs7$) \\
  1904 & H & 91/12 &  8303.045--303.079 &  2557 &  &                              
Djo\,1 & H & 94/09 &  9624.686--624.719 &  2428 & 20$'$ \\                        
  3201 & P & 91/12 &  8602.544--603.804 &  1787 &  &                              
Ter\,6 & H & 92/03 &  8697.825--699.179 &  4997 & \\                              
  4372 & P & 93/09 &  9234.920--237.030 &  8572 &  &                              
  6453 & H & 92/03 &  8696.966--702.298 &  2755 & \\                              
  4590 & H & 93/09 &  9729.609--731.504 &  6999 &  &                              
  6453 & P & 93/03 &  9069.165--072.306 & 28760 & 33$'$ \\                        
  5053 & H & 95/06 &  9890.716--894.914 &  6299 &  &                              
  6496 & P & 93/03 &  9077.465--077.643 & 11569 & \\                              
  5272 & H & 92/01 &  8633.160--633.182 &  1881 &  &                              
  6522 & P & 93/03 &  9076.031--076.947 & 29578 & $3\farcm4$\\                    
  5272 & H & 95/07 &  9913.062--913.162 &  4756 & $-0\fs11,-2\farcs1$ &           
  6528 & P & 93/03 &  9076.031--076.947 & 29578 & 19$'$\\                         
  5286 & H & 92/07 &  8835.408--836.422 &  5824 &  &                              
  6540 & P & 93/03 &  9078.463--078.704 & 10033 & 32$'$ \\                        
  5466 & P & 91/07 &  8450.203--451.079 & 25728 & 20$'$ &                         
  6541 & H & 91/03 &  8336.428--338.630 &  3837 & $-0\fs19,-1\farcs6$  \\         
  5466 & H & 94/07 &  9550.408--552.548 &  8079 &  &                              
  6544 & P & 91/03 &  8322.782--327.047 &  1686 & \\                              
  5466 & H & 95/01 &  9736.148--740.275 &  4041 & +3,$-$5  &                      
  6553 & P & 93/09 &  9241.314--241.338 &  2014 & 32$'$ \\                        
  5466 & H & 96/01 & 10090.270--098.387 &  4906 & +2,$-$2  &                      
Ter\,11& P & 93/09 &  9241.036--243.457 &  4616 & 24$'$ \\                        
  5824 & H & 91/02 &  8298.149--298.177 &  2432 &   &                             
  6626 & P & 91/03 &  8327.308--331.797 &  3334 & \\                              
  5824 & H & 94/08 &  9593.025--595.291 &  2122 &   &                             
  6626 & H & 95/03 &  9795.030--812.234 & 41764 & $-0\fs09,+0\farcs8$ \\          
  5904 & P & 93/07 &  9194.477--195.282 &  2017 &  &                              
  6626 & H & 96/03 & 10162.353--169.600 &  4573 & $+8,-4$  \\                     
  5904 & H & 94/02 &  9387.101--387.192 &  3574 &  &                              
  6626 & H & 96/09 & 10349.326--354.069 & 25541 & $-5,+6$ \\                      
  5904 & H & 94/08 &  9577.302--580.912 & 17969 & +14,$-$6  &                     
  6626 & H & 96/09 & 10355.244--355.799 &  8736 & $-6,+6$ \\                      
  5904 & H & 96/02 & 10132.222--134.227 &  3185 &  &                              
  6638 & P & 92/04 &  8716.730--721.395 & 16193 & 35$'$ \\                        
  5904 & H & 96/07 & 10288.351--320.863 & 28045 & +$0\fs17,-4\farcs8$ &           
  6638 & P & 93/09 &  9244.767--245.319 &  5981 & 41$'$ \\                        
  5986 & H & 95/09 &  9963.730--964.553 &  1613 & 19$'$  &                        
  6642 & P & 92/04 &  8714.671--718.869 &  7428 & \\                              
  6093 & P & 92/03 &  8691.506--691.733 &  9201 & $-0\fs03,0\farcs9$  &           
  6642 & P & 92/10 &  8904.672--908.262 & 20519 & 38$'$  \\                       
  6121 & P & 91/03 &  8317.311--317.390 &  2343 &  &                              
  6656 & P & 91/03 &  8329.713--330.329 &  8295 & \\                              
  6121 & P & 91/09 &  8510.101--510.229 &  2348 &  &                              
  6656 & H & 92/09 &  8881.759--882.975 &  9475 &  \\                             
  6121 & H & 93/02 &  9036.617--038.955 &  1584 &  &                              
  6656 & H & 93/03 &  9064.711--067.866 & 31709 &  \\                             
  6121 & H & 94/09 &  9595.829--612.044 & 12986 & +3,+2  &                        
  6715 & P & 91/03 &  8346.641--347.324 &  7871 & 41$'$ \\                        
  6121 & H & 95/08 &  9946.323--947.279 & 17484 & +$0\fs14,-1\farcs0$ &           
  6723 & P & 91/04 &  8348.571--349.412 &  7761 & 23$'$ \\                        
  6139 & H & 92/09 &  8881.472--881.487 &  1261 &  &                              
  6723 & H & 94/09 &  9626.344--627.839 &  6994 & \\                              
  6139 & H & 94/08 &  9588.787--615.527 &  9300 &  &                              
  6723 & H & 94/10 &  9633.244--633.285 &  3500 & $+3,0$\\                        
  6139 & P & 92/02 &  8679.257--679.278 &  1857 & 53$'$  &                        
  6760 & P & 92/03 &  8705.738--705.879 &  1548 & \\                              
  6139 & P & 93/03 &  9052.507--052.522 &  1316 & 53$'$  &                        
  6760 & P & 92/10 &  8911.056--912.727 & 14355 & $0,0$\\                         
  6205 & P & 92/09 &  8869.968--872.916 & 45872 &   &                             
  6760 & P & 93/03 &  9070.700--072.923 & 12544 & 27$'$ \\                        
  6205 & H & 94/09 &  9600.051--600.999 & 21634 &   &                             
  6779 & H & 98/03 & 10895.906--926.498 & 21169 & \\                              
  6254 & P & 93/09 &  9232.327--232.749 &  9420 &  &                              
  6809 & P & 93/03 &  9074.440--093.176 & 19952 & \\                              
  6266 & P & 91/03 &  8329.239--333.260 & 13868 & 15$'$  &                        
  6809 & H & 96/10 & 10378.463--385.441 &  5747 & \\                              
  6266 & P & 92/02 &  8680.558--681.031 & 12516 & 0,+11  &                        
  6809 & H & 97/09 & 10721.429--734.849 & 45561 & \\                              
  6273 & H & 92/03 &  8694.633--695.192 &  5971 &  &                              
  6838 & P & 91/04 &  8351.853--352.608 &  7492 & $21'$\\                         
  6293 & P & 92/09 &  8887.279--887.290 & 963 & $53'$ &                                
  6838 & H & 94/04 &  9452.860--466.812 & 31993 &  \\                             
  6293 & P & 92/09 &  8888.275--888.287 & 923 & $53'$ &                                
  6838 & H & 94/10 &  9635.799--637.477 & 27966 & $-9,+4$ \\                      
  6304 & H & 92/03 &  8695.631--696.255 &  5060 &  &                              
  7089 & H & 94/11 &  9685.045--688.126 &  8729 & \\                              
  6316 & H & 92/03 &  8695.697--696.840 &  6475 &  &                              
  7099 & P & 91/05 &  8386.592--386.607 &  1291 & $-9,-13$ \\                     
  6341 & H & 94/04 &  9447.410--450.952 & 18117 & $+1,-4$  &                      
  7099 & P & 91/11 &  8576.471--577.554 &  3000 & \\                              
  6341 & H & 94/08 &  9593.840--599.074 & 20254 & $+0\fs16,-0\farcs2$ &           
  7099 & P & 93/11 &  9299.339--299.759 & 10463 & 29$'$ \\                        
  6341 & H & 95/04 &  9817.436--817.468 &  2718 &  &                              
  7492 & P & 93/11 &  9320.934--321.151 &  4529 & 12$'$  \\                       
  6341 & P & 92/04 &  8709.449--709.730 &  6646 &  \\
\end{tabular}
\end{table*}

Starting from the list of globular clusters belonging to our Galaxy
with their positions as given by Djorgovski \&\ Meylan (1993), I use a 
ROSAT database utility to identify all ROSAT PSPC and HRI observations that 
have a globular cluster in the field of view. 
All observations thus found are listed in Table\,\ref{tobse} and
are analysed for this paper; with the exception of the HRI observations of
NGC\,104, NGC\,5139, NGC\,6397, NGC\,6440, NGC\,6752 and Liller\,1,
which have been reported elsewhere (Verbunt \&\ Hasinger
1998, Verbunt \&\ Johnston 2000, Verbunt et al.\ 2000), as well as
some data on NGC\,5272, published by Dotani et al.\ (1999).
The analysis of the X-ray data for each cluster is done in a 
number of steps.

The first of these steps uses the {\em standard reduction} with EXSAS
(Zimmermann et al.\ 1998) to find all detected sources in each
separate observation.  The standard analysis for HRI observations is
done as described in Verbunt \&\ Johnston (2000). The same method is
used for PSPC observations, except that for these no pixel size
correction is needed; to reduce background, the analysis -- unless
otherwise indicated -- discussed in this paper is limited to the
energy channels 50-240, i.e.\ energy range 0.5-2.5\,keV.  The standard
reduction produces for each source a maximum-likelihood value ML such
that the probability that the source is due to chance at one trial
position is $e^{-\rm ML}$. I retain for further discussion all sources
that have ML$\geq$$13$.

If more than one observation with the same detector was made, the {\em
offset between} these {\em observations} is determined, and the
observations are added, with the method described in Verbunt \&\
Hasinger (1998).  The determined offsets are rounded off to the
nearest integer number of pixels (1 pixel corresponding to
$0\farcs5$), and are given in Table\,\ref{tobse}.  The co-added image
is analysed with the standard procedure.

When no source is found in the cluster core, an {\em upper limit} to the
countrate is determined. For the PSPC I use the EXSAS procedure. This
uses a fit made to the spatial distribution of counts that remains after
removal of the detected sources. This fit predicts the number of background
counts at the required position, and sets the upper limit to the source counts
at the $n$-$\sigma$ excess above the map. I use $n$$=$2. Dividing this number 
of counts by the effective exposure time gives the upper limit to the 
countrate.
For the HRI the number of counts predicted at the source position from the 
background map is usually too small for this method to be reliable, and thus
I use the method described in Verbunt \&\ Johnston (2000) for HRI upper limits.
When no source is detected in or near the globular cluster, the
determination of the upper limit ends my analysis of that cluster.

When the cluster core contains a source which the standard analysis
labels as an {\em extended source}, a further analysis is used to separate it
into individual point sources. The method of this further analysis
is as described in Sect.\,2.3 of Verbunt et al.\ (2000). 
This method produces for each source a $\Delta\chi^2$ which 
may be compared with a $\chi^2$ distribution for three degrees of freedom.
Thus, a source with $\Delta\chi^2>14.2$ has a significance of more than 
3-$\sigma$. This statistical measure is correct only if the point
spread function is entirely accurate; to allow for uncertainty in the point
spread function, especially in longer exposures where observations of many 
separate orbits are added, I use 20 as the lower limit for $\Delta\chi^2$
at which a source is accepted.

A cluster source for which more than about 250 counts are detected
with the PSPC is subjected to {\em spectral analysis}. I extract source plus 
background counts from a circle around the source position; and background
from a ring around that. The source counts are found for each of the
channels 11-240 by subtracting the background in the circle, as estimated 
from the ring. The circle size is adjusted so that the net source counts
for all channels together agrees with the number found from the source 
detection algorithm.
I then bin the source counts, demanding that each bin has a minimum width 
equal to the spectral resolution and contains at least 20 photons.
This allows $\chi^2$ fitting. For this purpose the resolution $n_{\rm min}$ in
number of channels as a function of the channel number $n$ is sufficiently
well approximated with $n_{\rm min}=8+0.08n$, rounded off to nearest integer.
First, I try fits with a fixed absorption column $\nh$, derived from the
visual absorption $A_{\rm V}$ according to the empirical relation determined
by Predehl \&\ Schmitt (1995), $\nh=1.79\times10^{21}A_{\rm V}$\,\cmsq.
\nocite{ps95} 
If no satisfactory fit is found, the absorption is also allowed to vary,
increasing by one of the number of fitted parameters.
I approximate a bremsstrahlung spectrum with energy flux 
$f_{\nu}\propto\nu^{-0.3}\exp(-h\nu/kT)$.

The main goal of the identification of X-ray sources not related to
the globular cluster is an accurate determination
of the offset between the X-ray and optical coordinate systems
i.e.\ the bore sight correction.
For this reason I limit the search for {\em optical identifications}
to X-ray sources with positional accuracy (in the X-ray coordinate
frame) better than 10$''$.
I use SIMBAD to list objects within 1$'$ of each X-ray position;
possible identifications are well within this limit, always
within 10$''$ of the X-ray position, and most often rather
better than that.
Each suggested identification with a star is subjected to a
further check whether the observed countrate is within the
range expected for that star; the results are listed in the
Appendix, Table\,\ref{tstar}.
(I define this expected range on the basis of detections of nearby
stars in the ROSAT All Sky Survey, as explained in the Appendix.)
The offset between the X-ray and optical coordinate frames is
determined from these identified X-ray sources, and used to bring
the X-ray positions to the J2000 system.   
These offsets are also given in Table\,\ref{tobse}.

The probability of cluster membership as opposed to chance positional 
coincidence, is discussed after collation of the results for all
analyzed observations, in Sect.\,\ref{secprob}.

\section{Individual clusters}

In this section I describe the detected sources for the individual clusters,
which are listed in Table\,\ref{tsour}.
To limit the length of this Table, I tabulate sources only for
those clusters in or near which a (possibly) associated source is detected;
and only those non-cluster sources with positional accuracy better than 10$''$.
The luminosities of the cluster sources are listed in Table\,\ref{tsum},
which includes upper limits, for sources in the core and in Table\,\ref{tlist} 
for sources outside the cores.

\subsection{NGC\,104/47\,Tuc}

The ROSAT HRI observations of 47\,Tuc are described by Verbunt \&\ Hasinger 
(1998), \nocite{vh98}
who detect five sources in the cluster core, and four outside the core but
within the half-mass radius; all probably related to 47\,Tuc.
Reanalysis of the sources in the core with the multi-source fits described in 
Verbunt et al.\ (2000) gives the same positions and countrates as given in 
Table\,2 of Verbunt \&\ Hasinger, within the errors.
The source listed in Tab.\,\ref{tsum} is the conglomerate of
X\,5, X\,7, X\,9, X\,10 and X\,19 of Verbunt \&\ Hasinger.
\nocite{vh98}

The PSPC observations of 47\,Tuc have a total exposure of 59.8 ksec; as they
do not resolve the core, I have limited my analysis to the determination
in the two longer observations (listed in Table\,\ref{tobse})
of the countrate and of a fit to the spectrum of the conglomerate 
source in the cluster center. The data given in Table\,\ref{tsum} and
in Fig.\,\ref{fspec} are from the 1992 observation; the data from 1991 are
compatible with these, i.e.\ no indication of variability is found.
A bremsstrahlung spectrum absorbed by a column about twice that expected
on the basis of the visual absorption gives the best fit, which however
is still not acceptable (see Fig.\,\ref{fspec}). 
This may simply reflect that the
observed spectrum is the sum of several, non-identical spectra;
alternatively, bremsstrahlung may not correctly describe the spectra.
Blackbody or powerlaw spectra give even worse fits.
From the summed spectrum it is clear, however, that the flux from
the core is soft, peaking at an energy of about 0.6\,keV.

\begin{table}[]
\caption[o]{For each cluster I list the sources found by the standard 
EXSAS analysis,
and where applicable the sources found with a multiple-source
algorithm,
labeled by adding a lower-case letter to the number of the extended source
from the standard analysis.
For each cluster the core radius $r_c$ and half-mass radius $r_h$ are
given.
Sources are marked $\bullet$ when related to the cluster, 
$\circ$ when probably related to the cluster; for these sources x 
indicates that they are extended.
For each source I give the position, with error $\Delta$ in $''$,
an indication ML/$\Delta\chi^2$ of the source significance (explained in 
Sect.\,2, values larger than 99 are given as 99), the distance $d$ to 
the cluster center (in $r_c$), and the countrate. 
When both PSPC and HRI observations are available, the listed sources are
from the HRI observation; when more than one HRI observation is available,
the listed sources are from the added image.
The error in the position is the statistical error 
on the detector; to obtain the overall error the projection error must be
added in quadrature. The projection error is $\sim5''$ in the absence of
secure optical identifications; when secure optical identifications have been
made (as indicated by the shifts in arcseconds listed with the observation
date), the projection error is given in the text.
\label{tsour}}
\begin{tabular}{rr@{ }r@{ }rr@{ }r@{ }rc@{ }r@{ }r@{ }r}
X  &  \multicolumn{3}{c}{$\alpha$\,(2000)}
   &  \multicolumn{3}{c}{$\delta$\,(2000)} & $\Delta$ & ML & $d$ &
cts/ksec  \\ 
\multicolumn{11}{l}{{\bf  NGC\,362 } $r_c=11\farcs4, r_h/r_c=4.3$} \\
$\bullet$x1 & 01 & 03 & 14.95 & $-$70 & 50 & 57.9 & 1.2 &  44 &  0.4&  2.8$\pm$0.4 \\
          2 & 01 & 03 & 26.24 & $-$70 & 53 & 39.6 & 0.7 &  81 & 15.4&  2.5$\pm$0.4 \\
\multicolumn{7}{l}{NGC\,362 multi-source analysis}  & & $\Delta\chi^2$ \\
$\bullet$1a & 01 & 03 & 15.22 & $-$70 & 51 & 00.3 & 1.1 &  83 &  0.7&  1.7$\pm$0.3 \\
$\bullet$1b & 01 & 03 & 14.74 & $-$70 & 50 & 51.4 & 1.2 &  32 &  0.3&  1.0$\pm$0.3 \\
\multicolumn{8}{l}{{\bf  Pal\,2   } $r_c=14\farcs4, r_h/r_c=2.8$} & ML \\
   $\circ$1 & 04 & 46 & 03.85 &    31 & 22 & 38.1 & 1.4 &  16 &  2.0&  1.5$\pm$0.5 \\ 
\multicolumn{11}{l}{{\bf  NGC\,1904} $r_c=9\farcs6, r_h/r_c=5.0$} \\
$\circ$1    & 05 & 24 & 15.67 & $-$24 & 31 & 12.1 & 1.3 &  24 &  7.4&  4.1$\pm$1.3 \\
\multicolumn{11}{l}{{\bf  NGC\,5272/M\,3} $r_c=33\farcs0, r_h/r_c=2.0$} \\
          B & 13 & 42 & 10.86 &    28 & 28 & 47.8 & 0.8 & 99 & 11.4& 11.7$\pm$1.6 \\ 
          C & 13 & 42 & 54.47 &    28 & 28 & 04.3 & 1.8 & 99 & 20.0& 15.1$\pm$2.0 \\ 
 $\bullet$A & 13 & 42 & 09.76 &    28 & 22 & 47.2 & 1.1 &  55 &  0.7&  5.8$\pm$1.2 \\ 
          E & 13 & 41 & 15.43 &    28 & 16 & 04.9 & 3.7 &  40 & 25.2& 10.4$\pm$1.8 \\ 
          G & 13 & 42 & 28.44 &    28 & 14 & 29.5 & 2.4 &  19 & 16.2&  3.4$\pm$0.9 \\ 
O  &  \multicolumn{3}{c}{$\alpha$\,(2000)}
   &  \multicolumn{3}{c}{$\delta$\,(2000)} \\
B & 13 & 42 & 10.89 & 28 & 28 & 47.1 & \multicolumn{4}{l}{QSO Q\,1339+2843$^{a,b}$}\\
C & 13 & 42 & 54.41 & 28 & 28 & 06.5 & \multicolumn{4}{l}{QSO B2\,1340+28$^{a,c}$}\\
E & 13 & 41 & 15.31 & 28 & 16 & 04.4 & \multicolumn{4}{l}{QSO 1338+2831$^a$}\\
\multicolumn{10}{l}{$^a$\,Harris et al.\ 1992, $^b$\,Carney 1976,
$^c$\,Geffert 1998}\\
\multicolumn{11}{l}{{\bf NGC\,5904} $r_c=25\farcs2, r_h/r_c=5.0$} \\
          1 & 15 & 17 & 57.87 &    02 & 15 & 39.4 & 2.4 &  47 & 33.2&  2.4$\pm$0.3 \\ 
          2 & 15 & 18 & 24.14 &    02 & 13 & 46.6 & 1.2 &  67 & 21.7&  1.8$\pm$0.2 \\ 
          3 & 15 & 18 & 58.59 &    02 & 07 & 57.3 & 1.3 &  34 & 16.4&  0.9$\pm$0.2 \\ 
 $\bullet$4 & 15 & 18 & 14.47 &    02 & 05 & 35.3 & 1.4 &  14 & 11.6&  0.4$\pm$0.1 \\ 
          6 & 15 & 19 & 26.76 &    02 & 03 & 59.7 & 3.0 &  35 & 31.6&  2.1$\pm$0.3 \\ 
          7 & 15 & 18 & 13.81 &    02 & 02 & 40.8 & 1.3 &  17 & 13.1&  0.4$\pm$0.1 \\ 
          5 & 15 & 18 & 31.93 &    01 & 58 & 49.4 & 2.3 &  13 & 14.7&  0.7$\pm$0.2 \\ 
          8 & 15 & 18 & 51.79 &    01 & 57 & 57.0 & 2.1 &  14 & 19.8&  0.6$\pm$0.2 \\ 
          9 & 15 & 18 &  4.61 &    01 & 57 & 49.0 & 1.6 &  55 & 24.3&  1.8$\pm$0.2 \\ 
         10 & 15 & 18 & 46.66 &    01 & 56 & 56.5 & 1.3 &  49 & 20.6&  1.4$\pm$0.2 \\ 
O  &  \multicolumn{3}{c}{$\alpha$\,(2000)}
   &  \multicolumn{3}{c}{$\delta$\,(2000)} \\
4 & 15 & 18 & 14.46 & 02 & 05 & 35.2 & \multicolumn{4}{l}{V\,101$^a$} \\
6 & 15 & 19 & 26.8\phantom{0} & 02 & 04 & 00.0 & 
\multicolumn{4}{l}{QSO Q\,1516+0214$^b$}\\
\multicolumn{11}{l}{$^a$\,Evstigneeva et al.\ 1995, $^b$\,Harris et al.\ 1992}\\
\end{tabular}
\end{table}
\setcounter{table}{1}
\begin{table}
\caption{Continued}
\begin{tabular}{rr@{ }r@{ }rr@{ }r@{ }rc@{ }r@{ }r@{ }r}
X  &  \multicolumn{3}{c}{$\alpha$\,(2000)}
   &  \multicolumn{3}{c}{$\delta$\,(2000)} & $\Delta$ & ML & $d$ &
cts/ksec  \\ 
\multicolumn{11}{l}{{\bf NGC\,6093/M\,80} $r_c=9\farcs0, r_h/r_c=4.3$}\\
          1 & 16 & 15 & 34.54 & $-$22 & 42 & 41.4 & 1.9 &  99 &171.2&178.7$\pm$5.1 \\ 
          2 & 16 & 16 & 12.34 & $-$22 & 44 & 24.8 & 7.7 &  47 &121.4&  7.1$\pm$1.1 \\ 
          3 & 16 & 17 & 14.13 & $-$22 & 51 & 56.6 & 4.6 &  27 & 47.2&  2.3$\pm$0.6 \\ 
          4 & 16 & 16 & 37.53 & $-$22 & 52 & 03.6 & 5.2 &  18 & 57.5&  1.7$\pm$0.5 \\ 
          5 & 16 & 17 & 31.07 & $-$22 & 52 & 14.1 & 1.9 &  99 & 60.6& 37.4$\pm$2.1 \\ 
          6 & 16 & 17 & 12.38 & $-$22 & 53 & 08.8 & 4.9 &  20 & 38.8&  1.7$\pm$0.5 \\ 
          7 & 16 & 17 & 14.67 & $-$22 & 55 & 19.8 & 1.5 &  99 & 28.2& 13.9$\pm$1.3 \\ 
          8 & 16 & 17 & 09.24 & $-$22 & 55 & 22.0 & 6.3 &  16 & 23.3&  1.8$\pm$0.5 \\ 
          9 & 16 & 16 & 46.96 & $-$22 & 57 & 34.5 & 5.2 &  14 & 24.6&  1.4$\pm$0.5 \\ 
$\bullet$10 & 16 & 17 & 02.27 & $-$22 & 58 & 30.0 & 3.7 &  45 &  0.4&  3.3$\pm$0.7 \\ 
         11 & 16 & 16 & 35.09 & $-$23 & 02 & 21.3 & 5.3 &  17 & 49.3&  1.9$\pm$0.5 \\ 
         12 & 16 & 17 & 53.16 & $-$23 & 02 & 29.1 & 3.5 &  68 & 82.2&  5.1$\pm$0.8 \\ 
         13 & 16 & 17 & 11.59 & $-$23 & 03 & 03.0 & 5.2 &  21 & 33.4&  1.9$\pm$0.5 \\ 
         14 & 16 & 17 & 31.20 & $-$23 & 03 & 32.8 & 0.2 &  99 & 55.4&550.7$\pm$7.9 \\ 
         15 & 16 & 14 & 12.12 & $-$23 & 04 & 56.7 & 8.6 &  99 &264.9&163.5$\pm$6.1 \\ 
         16 & 16 & 16 & 39.31 & $-$23 & 10 & 02.8 & 7.9 &  15 & 84.8&  2.1$\pm$ 0.6 \\
         17 & 16 & 16 & 40.79 & $-$23 & 14 & 35.4 & 4.5 &  99 &112.3&  9.6$\pm$ 1.2 \\
         18 & 16 & 16 & 00.21 & $-$23 & 25 & 10.2 & 6.6 &  99 &201.9& 42.1$\pm$ 2.9 \\
O  &  \multicolumn{3}{c}{$\alpha$\,(2000)}
   &  \multicolumn{3}{c}{$\delta$\,(2000)} \\
 1 & 16 & 15 & 34.72 & $-$22 & 42 & 38.0 & \multicolumn{3}{l}{VV\,Sco} \\
 7 & 16 & 17 & 14.67 & $-$22 & 55 & 19.8 & \multicolumn{3}{l}{HD\,146457}\\
14 & 16 & 17 & 31.41 & $-$23 & 03 & 36.1 & \multicolumn{3}{l}{HD\,146516}\\
\multicolumn{11}{l}{{\bf NGC\,6121/M\,4} $r_c=49\farcs8, r_h/r_c=4.4$}\\
          7 & 16 & 23 & 57.30 & $-$26 & 20 & 27.2 & 3.2 &  43 & 14.6&  3.4$\pm$ 0.5 \\
          3 & 16 & 23 & 22.93 & $-$26 & 22 & 16.4 & 0.2 &  99 & 11.6& 58.5$\pm$ 1.4 \\
          8 & 16 & 24 & 14.93 & $-$26 & 27 & 52.7 & 2.6 &  15 & 11.5&  1.0$\pm$ 0.3 \\
 $\bullet$9 & 16 & 23 & 34.24 & $-$26 & 31 & 41.8 & 1.7 &  11 &  0.4&  0.5$\pm$ 0.2 \\
         10 & 16 & 24 & 26.97 & $-$26 & 35 & 33.3 & 4.4 &  14 & 14.7&  1.5$\pm$ 0.4 \\
O  &  \multicolumn{3}{c}{$\alpha$\,(2000)}
   &  \multicolumn{3}{c}{$\delta$\,(2000)} \\
 3 & 16 & 23 & 22.93 & $-$26 & 22 & 16.4 & 
\multicolumn{3}{l}{HIP\,80290$^a$} \\
\multicolumn{10}{l}{$^a$\,Perryman et al.\ 1997}\\
\multicolumn{11}{l}{{\bf  NGC\,6139} $r_c=8\farcs4, r_h/r_c=5.9$}\\
 $\bullet$1 & 16 & 27 & 40.58 & $-$38 & 50 & 56.2 & 1.6 &  20 &  0.3&  1.4$\pm$ 0.4 \\
          2 & 16 & 27 & 41.84 & $-$38 & 53 & 22.4 & 1.3 &  29 & 17.5&  2.1$\pm$ 0.5 \\
\multicolumn{11}{l}{{\bf NGC\,6205/M\,13} $r_c=$ 46.8$'', r_h/r_c=$  1.9}\\
          N & 16 & 41 & 53.13 &    36 & 40 & 21.7 & 5.0 &  14 & 16.6&  2.4$\pm$ 0.5 \\
          R & 16 & 40 & 47.17 &    36 & 37 & 51.8 & 4.5 &  15 & 19.2&  2.5$\pm$ 0.6 \\
          H & 16 & 41 & 56.79 &    36 & 35 & 12.9 & 2.5 &  17 & 10.5&  1.4$\pm$ 0.3 \\
          B & 16 & 40 & 56.42 &    36 & 34 & 05.3 & 0.8 &  99 & 14.3&  8.7$\pm$ 0.7 \\
          D & 16 & 41 & 11.17 &    36 & 32 & 30.8 & 0.7 &  99 & 10.0&  5.2$\pm$ 0.5 \\
          J & 16 & 42 & 19.87 &    36 & 31 & 50.4 & 1.6 &  34 & 11.3&  2.0$\pm$ 0.4 \\
 $\bullet$xG & 16 & 41 & 44.02 &    36 & 27 & 58.0 & 1.0 &  51 &  0.8&  2.0$\pm$ 0.3 \\
          E & 16 & 41 & 18.65 &    36 & 26 & 47.2 & 1.2 &  20 &  6.0&  0.9$\pm$ 0.2 \\
   $\circ$F & 16 & 41 & 38.47 &    36 & 26 & 28.7 & 3.6 &  99 &  1.6&  2.4$\pm$0.2$^a$\\
          C & 16 & 40 & 58.77 &    36 & 24 & 27.6 & 1.7 &  31 & 11.7&  1.9$\pm$ 0.4 \\
          I & 16 & 42 & 03.62 &    36 & 24 & 14.1 & 1.3 &  21 &  7.2&  1.0$\pm$ 0.3 \\
          A & 16 & 40 & 32.13 &    36 & 23 & 19.7 & 4.6 &  16 & 18.7&  2.6$\pm$ 0.6 \\
          Q & 16 & 42 & 35.66 &    36 & 18 & 10.8 & 5.8 &  10 & 18.5 & 2.2$\pm$0.6 \\
          S & 16 & 40 & 55.85 &    36 & 16 & 01.4 & 6.1 &  11 & 18.9 & 2.3$\pm$0.6 \\
          T & 16 & 41 & 18.08 &    35 & 59 & 32.7 & 4.4 &  99 & 36.5 & 19.4$\pm$0.8$^a$ \\
\multicolumn{7}{l}{NGC\,6205 multi-source analysis}  & & $\Delta\chi^2$ \\
$\bullet$Ga & 16 & 41 & 44.02 & 36 & 27 & 57.7 & 0.9 &  70 & 0.8 &  1.8$\pm$0.3 \\
$\bullet$Gb & 16 & 41 & 45.10 & 36 & 27 & 35.5 & 2.7 &  21 & 0.9 &  0.7$\pm$0.2 \\
O  &  \multicolumn{3}{c}{$\alpha$\,(2000)}
   &  \multicolumn{3}{c}{$\delta$\,(2000)} \\
 R & 16 & 40 & 46.83 & 36 & 37 & 53.51 & 
\multicolumn{4}{l}{USNO-A2\,1200-07982833} \\
 T & 16 & 41 & 17.53 & 35 & 59 & 31.26 &
\multicolumn{4}{l}{USNO-A2\,1200-07987402} \\
\multicolumn{10}{l}{$^a$\,countrate for PSPC channels 50-240} \\
\end{tabular}
\end{table}
\setcounter{table}{1}
\begin{table}
\caption{Continued}
\begin{tabular}{rr@{ }r@{ }rr@{ }r@{ }rc@{ }r@{ }r@{ }r}
X  &  \multicolumn{3}{c}{$\alpha$\,(2000)}
   &  \multicolumn{3}{c}{$\delta$\,(2000)} & $\Delta$ & ML & $d$ &
cts/ksec  \\ 
\multicolumn{11}{l}{{\bf  NGC\,6266/M\,62} $r_c=10\farcs8, r_h/r_c=6.8$}\\
          1 & 17 & 02 & 04.90 & $-$29 & 42 & 59.9 & 6.5 &  29 &146.1&  2.6$\pm$ 0.5 \\
          2 & 17 & 03 & 24.32 & $-$29 & 48 & 42.8 & 6.9 &  46 &187.3&  4.8$\pm$ 0.6 \\
          3 & 17 & 02 & 04.66 & $-$29 & 48 & 58.0 & 4.9 &  14 &116.9&  1.0$\pm$ 0.3 \\
          4 & 17 & 00 & 27.79 & $-$29 & 49 & 49.6 & 8.2 &  29 &108.3&  3.7$\pm$ 0.6 \\
          5 & 17 & 02 & 44.82 & $-$29 & 50 & 14.7 & 4.2 &  32 &143.8&  2.0$\pm$ 0.4 \\
          6 & 17 & 03 & 07.29 & $-$29 & 50 & 27.1 & 1.1 &  99 &164.8& 38.7$\pm$ 1.3 \\
          7 & 17 & 01 & 59.17 & $-$29 & 50 & 28.6 & 4.9 &  17 &106.2&  1.2$\pm$ 0.3 \\
          8 & 17 & 01 & 10.02 & $-$29 & 54 & 41.1 & 1.9 &  99 & 67.0&  6.9$\pm$ 0.6 \\
          9 & 17 & 01 & 28.82 & $-$29 & 57 & 46.4 & 4.2 &  26 & 53.5&  1.6$\pm$ 0.3 \\
         10 & 17 & 00 & 50.64 & $-$29 & 58 & 41.0 & 3.0 &  99 & 51.9&  7.4$\pm$ 0.7 \\
         11 & 17 & 02 & 08.51 & $-$30 & 03 &  6.7 & 3.9 &  25 & 70.1&  1.5$\pm$ 0.3 \\
         12 & 17 & 00 & 16.54 & $-$30 & 03 & 32.3 & 1.3 &  99 & 69.7&132.7$\pm$ 2.6 \\
         13 & 17 & 03 & 06.41 & $-$30 & 06 & 17.8 & 7.3 &  17 &136.8&  1.8$\pm$ 0.4 \\
$\bullet$x14 & 17 & 01 & 13.08 & $-$30 & 06 & 44.3 & 1.9 &  99 &  0.6& 22.1$\pm$ 1.1 \\
         15 & 17 & 02 & 10.57 & $-$30 & 14 &  8.9 & 8.2 &  18 & 80.9&  1.9$\pm$ 0.4 \\
O  &  \multicolumn{3}{c}{$\alpha$\,(2000)}
   &  \multicolumn{3}{c}{$\delta$\,(2000)} \\
-- & 16 & 58 & 55.4 & $-$29 & 52 & 28 & 
\multicolumn{4}{l}{MXB\,1658$-$298$^a$} \\
8 & 17 & 01 & 09.81 & $-$29 & 54 & 42.0 & \multicolumn{4}{l}{PKS\,1657$-$28$^b$} \\
12 & 17 & 00 & 16.59 & $-$30 & 03 & 32.1 & \multicolumn{4}{l}{TYC\,7360\,394\,1$^c$}\\
\multicolumn{10}{l}{$^a$\,Cominsky \&\ Wood 1989, $^b$\,Wright et al.\ 1991}\\
\multicolumn{10}{l}{$^c$\,H{\o}g et al.\ 1997}\\
\multicolumn{11}{l}{{\bf  NGC\,6341/M\,92} $r_c=13\farcs8, r_h/r_c=4.7$}\\
          2 & 17 & 17 & 30.28 &    43 & 19 & 43.8 & 3.1 &  18 & 53.4&  1.5$\pm$ 0.3 \\
          E & 17 & 16 & 38.17 &    43 & 12 & 15.7 & 1.8 &  15 & 29.1&  0.6$\pm$ 0.2 \\
          6 & 17 & 17 & 01.26 &    43 & 10 & 43.6 & 0.9 &  55 & 12.1&  1.3$\pm$ 0.2 \\
          7 & 17 & 18 & 10.61 &    43 & 08 & 56.8 & 2.8 &  22 & 50.3&  1.5$\pm$ 0.3 \\
 $\bullet$8 & 17 & 17 & 07.19 &    43 & 08 & 14.0 & 1.2 &  46 &  0.2&  1.4$\pm$ 0.2 \\
         10 & 17 & 16 & 32.72 &    43 & 02 & 29.1 & 0.2 &  99 & 37.0& 29.3$\pm$ 0.9 \\
         12 & 17 & 16 & 49.29 &    42 & 54 & 55.3 & 3.1 &  24 & 59.4&  2.0$\pm$ 0.4 \\
O  &  \multicolumn{3}{c}{$\alpha$\,(2000)}
   &  \multicolumn{3}{c}{$\delta$\,(2000)} \\
E & 17 & 16 & 38.27 & 43 & 12 & 14.4 & \multicolumn{4}{l}{V798\,Her$^a$} \\
6 & 17 & 16 & 32.70 & 43 & 02 & 29.4 & \multicolumn{4}{l}{TYC\,3081\,510\,1$^b$} \\
\multicolumn{10}{l}{$^a$\,Tucholke et al.\, 1996, $^b$\,H{\o}g et al.\ 1997}\\
\multicolumn{11}{l}{{\bf  NGC\,6352} $r_c=49\farcs8, r_h/r_c=2.4$}\\
          2 & 17 & 24 & 32.14 & $-$48 & 16 & 18.4 & 5.1 &  36 & 15.8&  3.8$\pm$ 0.8 \\
          3 & 17 & 24 & 42.70 & $-$48 & 17 & 45.9 & 6.8 &  21 & 13.0&  2.8$\pm$ 0.7 \\
         17 & 17 & 25 & 17.72 & $-$48 & 21 & 12.2 & 6.0 &  14 &  5.5&  1.5$\pm$ 0.5 \\
   $\circ$4 & 17 & 25 & 21.43 & $-$48 & 26 & 09.7 & 5.4 &  19 &  1.8&  1.9$\pm$ 0.5 \\
          5 & 17 & 25 & 20.76 & $-$48 & 29 & 41.2 & 4.0 &  53 &  5.5&  4.7$\pm$ 0.8 \\
          6 & 17 & 26 & 34.40 & $-$48 & 30 & 44.6 & 5.2 &  26 & 14.6&  2.7$\pm$ 0.6 \\
          7 & 17 & 27 & 13.16 & $-$48 & 32 & 46.4 & 5.7 &  46 & 22.6&  5.5$\pm$ 0.9 \\
          8 & 17 & 27 & 50.90 & $-$48 & 32 & 49.4 & 7.2 &  98 & 29.7& 15.6$\pm$ 1.7 \\
          9 & 17 & 26 & 13.17 & $-$48 & 35 & 24.3 & 1.7 &  99 & 15.0& 20.2$\pm$ 1.6 \\
         10 & 17 & 26 & 22.51 & $-$48 & 36 & 54.9 & 3.0 &  99 & 17.5&  9.6$\pm$ 1.1 \\
         11 & 17 & 25 & 10.24 & $-$48 & 39 & 53.8 & 5.5 &  40 & 17.9&  4.3$\pm$ 0.8 \\
         12 & 17 & 24 & 40.66 & $-$48 & 40 & 07.2 & 5.2 &  71 & 20.3&  7.9$\pm$ 1.1 \\
         14 & 17 & 25 & 37.80 & $-$48 & 46 & 15.7 & 8.6 &  29 & 25.2&  4.8$\pm$ 1.0 \\
         16 & 17 & 24 & 34.44 & $-$48 & 59 & 50.4 & 9.4 &  99 & 43.0& 37.9$\pm$ 2.9 \\
O  &  \multicolumn{3}{c}{$\alpha$\,(2000)}
   &  \multicolumn{3}{c}{$\delta$\,(2000)} \\
9 & 17 & 26 & 13.22 & $-$48 & 35 & 25.4 & \multicolumn{4}{l}{HIP\,85326$^a$} \\
10 & 17 & 26 & 22.46 & $-$48 & 36 & 53.6 & \multicolumn{4}{l}{HIP\,85342$^a$}\\
\multicolumn{10}{l}{$^a$\,Perryman et al.\ 1997}\\
\end{tabular}
\end{table}
\setcounter{table}{1}
\begin{table}
\caption{Continued}
\begin{tabular}{rr@{ }r@{ }rr@{ }r@{ }rc@{ }r@{ }r@{ }r}
X  &  \multicolumn{3}{c}{$\alpha$\,(2000)}
   &  \multicolumn{3}{c}{$\delta$\,(2000)} & $\Delta$ & ML & $d$ &
cts/ksec  \\ 
\multicolumn{11}{l}{{\bf  NGC\,6366} $r_c=109\farcs8, r_h/r_c=1.4$}\\
          7 & 17 & 27 & 45.95 & $-$04 & 50 & 00.7 & 8.5 &  15 &  8.0&  1.8$\pm$ 0.5 \\
          8 & 17 & 27 & 44.25 & $-$04 & 54 & 18.6 & 5.3 &  25 &  5.6&  2.0$\pm$ 0.5 \\
          2 & 17 & 28 & 32.80 & $-$05 & 03 & 25.1 & 6.6 &  18 &  6.6&  1.6$\pm$ 0.5 \\
          3 & 17 & 26 & 52.40 & $-$05 & 03 & 57.5 & 8.8 &  15 &  7.1&  1.9$\pm$ 0.5 \\
 $\bullet$4 & 17 & 27 & 42.36 & $-$05 & 05 & 05.5 & 5.4 &  27 &  0.4&  2.4$\pm$ 0.6 \\
          5 & 17 & 26 & 37.88 & $-$05 & 05 & 11.7 & 1.7 &  99 &  9.1& 54.1$\pm$ 2.5 \\
          6 & 17 & 27 & 53.44 & $-$05 & 14 & 56.5 & 7.1 &  15 &  5.8&  1.5$\pm$ 0.5 \\
O  &  \multicolumn{3}{c}{$\alpha$\,(2000)}
   &  \multicolumn{3}{c}{$\delta$\,(2000)} \\
 5 & 17 & 26 & 37.88 & $-$05 & 05 & 11.7 & 
\multicolumn{4}{l}{HIP\,85365$^a$} \\
\multicolumn{10}{l}{$^a$\,Perryman et al.\ 1997} \\
\multicolumn{11}{l}{{\bf  NGC\,6388} $r_c=7\farcs2, r_h/r_c=5.6$}\\
   & \multicolumn{3}{c}{1991}\\
          1 & 17 & 38 & 06.02 & $-$44 & 03 & 32.5 & 1.5 &  99 &374.5&136.6$\pm$ 3.0 \\
          2 & 17 & 38 & 46.39 & $-$44 & 06 & 00.2 & 1.6 &  99 &386.9& 68.9$\pm$ 2.1 \\
          3 & 17 & 37 & 18.26 & $-$44 & 25 & 54.4 & 9.9 &  25 &176.7& 10.4$\pm$ 1.2 \\
          4 & 17 & 37 & 18.26 & $-$44 & 29 & 21.1 & 9.8 &  19 &152.7& 10.9$\pm$ 1.2 \\
   $\circ$5 & 17 & 36 & 20.44 & $-$44 & 44 & 39.9 & 9.3 &  92 &  6.9& 16.7$\pm$ 1.6 \\
   & \multicolumn{3}{c}{1992}\\
          1 & 17 & 38 & 06.08 & $-$44 & 03 & 33.8 & 4.2 & 99 &374.4& 95.5$\pm$ 9.1 \\
          6 & 17 & 38 & 40.34 & $-$44 & 17 & 40.5 & 3.5 &  99 &305.8&122.7$\pm$10.6 \\
          7 & 17 & 36 & 51.74 & $-$44 & 20 & 07.3 & 4.0 &  49 &206.3& 19.1$\pm$ 3.8 \\
   $\circ$5 & 17 & 36 & 17.30 & $-$44 & 44 & 52.7 &26.7 &  15 &  6.5& 29.7$\pm$ 6.8 \\
O  &  \multicolumn{3}{c}{$\alpha$\,(2000)}
   &  \multicolumn{3}{c}{$\delta$\,(2000)} \\
 1 & 17 & 38 & 06.10 & $-$44 & 03 & 31.6 &  
\multicolumn{4}{l}{TYC\,7896\,3885\,1$^a$} \\
 2 & 17 & 38 & 46.31 & $-$44 & 06 & 01.1 &
\multicolumn{4}{l}{HIP\,86356$^b$} \\
 6 & 17 & 38 & 40.37 & $-$44 & 17 & 42.8 &  
\multicolumn{4}{l}{TYC\,7896\,3812\,1$^a$} \\
 7 & 17 & 36 & 51.68 & $-$44 & 20 & 06.8 &  
\multicolumn{4}{l}{TYC\,7896\,2299\,1$^a$} \\
\multicolumn{10}{l}{$^a$\,H{\o}g et al.\ 1997, $^b$\,Perryman et al.\ 1997}\\
\multicolumn{11}{l}{{\bf  NGC\,6541} $r_c=18\farcs0, r_h/r_c=4.0$}\\
          1 & 18 & 08 & 41.41 & $-$43 & 36 & 26.4 & 1.2 &  99 & a & 18.1$\pm$ 2.3 \\
 $\bullet$2 & 18 & 08 & 01.74 & $-$43 & 42 & 57.0 & 1.2 &  54 & a &  6.6$\pm$ 1.4 \\
O  &  \multicolumn{3}{c}{$\alpha$\,(2000)}
   &  \multicolumn{3}{c}{$\delta$\,(2000)} \\
 1 & 18 &  8 & 41.41 & $-$43 & 36 & 26.4 &  
\multicolumn{4}{l}{TYC\,7911\,112\,1$^b$} \\
\multicolumn{10}{l}{$^a$center of cluster unknown, $^b$\,H{\o}g et al.\ 1997}\\
\multicolumn{11}{l}{{\bf  NGC\,6626} $r_c=14\farcs4, r_h/r_c=6.5$}\\
          5 & 18 & 24 & 07.27 & $-$24 & 47 & 19.1 & 1.5 &  18 & 31.6&  0.5$\pm$ 0.1 \\
          6 & 18 & 24 & 22.68 & $-$24 & 51 & 02.5 & 1.3 &  15 & 10.8&  0.3$\pm$ 0.1 \\
 $\bullet$2 & 18 & 24 & 32.73 & $-$24 & 52 & 08.8 & 0.3 &  99 &  0.3&  9.4$\pm$ 0.4 \\
   $\circ$7 & 18 & 24 & 31.01 & $-$24 & 52 & 46.3 & 1.0 &  34 &  3.0&  0.6$\pm$ 0.1 \\
          8 & 18 & 23 & 57.88 & $-$24 & 54 & 10.7 & 1.9 &  26 & 34.1&  0.7$\pm$ 0.1 \\
          9 & 18 & 24 & 24.27 & $-$24 & 58 & 07.6 & 2.0 &  19 & 26.0&  0.6$\pm$ 0.1 \\
         10 & 18 & 25 & 27.23 & $-$24 & 59 & 06.4 & 3.8 &  16 & 58.9&  1.1$\pm$ 0.2 \\
         11 & 18 & 24 & 27.61 & $-$24 & 59 & 10.4 & 1.7 &  15 & 29.5&  0.4$\pm$ 0.1 \\
         12 & 18 & 24 & 38.45 & $-$25 & 01 & 35.8 & 1.6 &  36 & 39.5&  1.0$\pm$ 0.2 \\
\multicolumn{9}{c}{X-ray sources near center; 3-source fit} \\
         2a & 18 & 24 & 32.99 & $-$24 & 52 & 10.1 & 0.7 &     &  0.2&  3.7$\pm$0.7 \\ 
         2b & 18 & 24 & 32.66 & $-$24 & 52 & 07.0 & 0.7 &     &  0.4&  4.0$\pm$0.7 \\ 
         2c & 18 & 24 & 32.15 & $-$24 & 52 & 11.1 & 0.7 &     &  0.7&  1.5$\pm$0.2 \\ 
O  &  \multicolumn{3}{c}{$\alpha$\,(2000)}
   &  \multicolumn{3}{c}{$\delta$\,(2000)} \\
 9 & 18 & 24 & 24.27 & $-$24 & 58 & 07.6 &  
\multicolumn{4}{l}{TYC\,6848\,3536\,1$^a$} \\
2c & 18 & 24 & 32.01 & $-$24 & 52 & 10.7 &
\multicolumn{4}{l}{PSR\,B1821-24$^b$} \\
\multicolumn{10}{l}{$^a$\,H{\o}g et al.\ 1997, $^b$\,Taylor et al.\ 1993}\\
\end{tabular}
\end{table}
\setcounter{table}{1}
\begin{table}
\caption{Continued}
\begin{tabular}{rr@{ }r@{ }rr@{ }r@{ }rc@{ }r@{ }rr}
X  &  \multicolumn{3}{c}{$\alpha$\,(2000)}
   &  \multicolumn{3}{c}{$\delta$\,(2000)} & $\Delta$ & ML & $d$ &
cts/ksec  \\ 
\multicolumn{11}{l}{{\bf  NGC\,6656} $r_c=85\farcs2, r_h/r_c=2.3$}\\
          3 & 18 & 35 & 50.90 & $-$23 & 46 & 49.9 & 2.8 &  48 &  7.5&  3.8$\pm$ 0.5 \\
          9 & 18 & 36 & 48.70 & $-$23 & 48 & 08.7 & 2.0 &  17 &  5.8&  0.9$\pm$ 0.2 \\
 $\bullet$4 & 18 & 36 & 24.79 & $-$23 & 54 & 35.6 & 0.8 &  36 &  0.3&  0.8$\pm$ 0.2 \\
\multicolumn{11}{l}{{\bf  NGC 6809} $r_c=169\farcs8, r_h/r_c=1.0$}\\
          8 & 19 & 40 & 42.45 & $-$30 & 56 & 28.6 & 1.3 &  57 &  3.3&  1.6$\pm$ 0.2 \\
 $\bullet$9 & 19 & 40 & 08.21 & $-$30 & 58 & 53.2 & 1.5 &  15 &  0.8&  0.5$\pm$ 0.1 \\
         11 & 19 & 40 & 43.70 & $-$30 & 59 & 03.3 & 2.1 &  21 &  3.4&  0.9$\pm$ 0.2 \\
         13 & 19 & 39 & 55.52 & $-$31 & 02 & 04.9 & 0.9 &  38 &  1.6&  0.7$\pm$ 0.1 \\
\multicolumn{11}{l}{{\bf  NGC 7099} $r_c=$  3\farcs6$, r_h/r_c=$ 19.2}\\
          1 & 21 & 41 &  4.71 & $-$22 & 50 & 27.7 & 4.2 &  99 &375.6& 58.3$\pm$ 4.3 \\
          3 & 21 & 39 & 23.47 & $-$22 & 55 & 53.3 & 8.1 &  39 &334.1&  8.8$\pm$ 1.7 \\
 $\bullet$8 & 21 & 40 & 22.39 & $-$23 & 10 & 41.9 & 2.9 &  90 &  1.7&  8.0$\pm$ 1.4 \\
         13 & 21 & 39 & 58.37 & $-$23 & 21 &  1.6 & 6.5 &  17 &193.7&  2.7$\pm$ 0.9 \\
          9 & 21 & 40 &  8.90 & $-$23 & 23 & 43.5 & 6.4 &  32 &222.0&  5.7$\pm$ 1.3 \\
          2 & 21 & 40 & 15.62 & $-$23 & 39 & 31.8 & 3.8 &  99 &480.3&168.5$\pm$ 7.5 \\
\end{tabular}
\end{table}

\subsection{NGC\,288}

The ROSAT HRI observation of this cluster was obtained on 1998 Jan 8-9,
and has been published by Sarazin et 
al.\ (1999), who discover a source with a countrate of 1.6$\pm$0.4 \ctks\ in
the core, at a distance of 12$''$ from the cluster center. (With the exposure
of about 20\,ks, this means that about 30 counts were detected.)
At the time of writing, the data are not public, so I have not reanalyzed
them.

\subsection{NGC\,362}

One HRI observation was made of NGC\,362, in which two sources are detected.
One of these, in the core of the cluster, is extended.
Further analysis of this source with the multi-source detection code described
in Sect.\,2 shows that the central source consists of two components.

\subsection{NGC\,1261}

NGC\,1261 lies in the field of view of the two PSPC observations of the 
Marano Field, albeit outside the area analyzed by Zamorani et al.\ (1999). 
\nocite{zmh+99}
In the longer of the PSPC observations no source at the cluster position is
found.

Two HRI observations of the cluster have been made. One source is detected
in the shorter observation; a source is detected at this
position in the longer observation with small significance. I have used
this source to determine the offset between the two observations, and
analyzed the combined image. Two sources, not related to the cluster,
were detected.

\subsection{Pal\,2}

\begin{figure*}
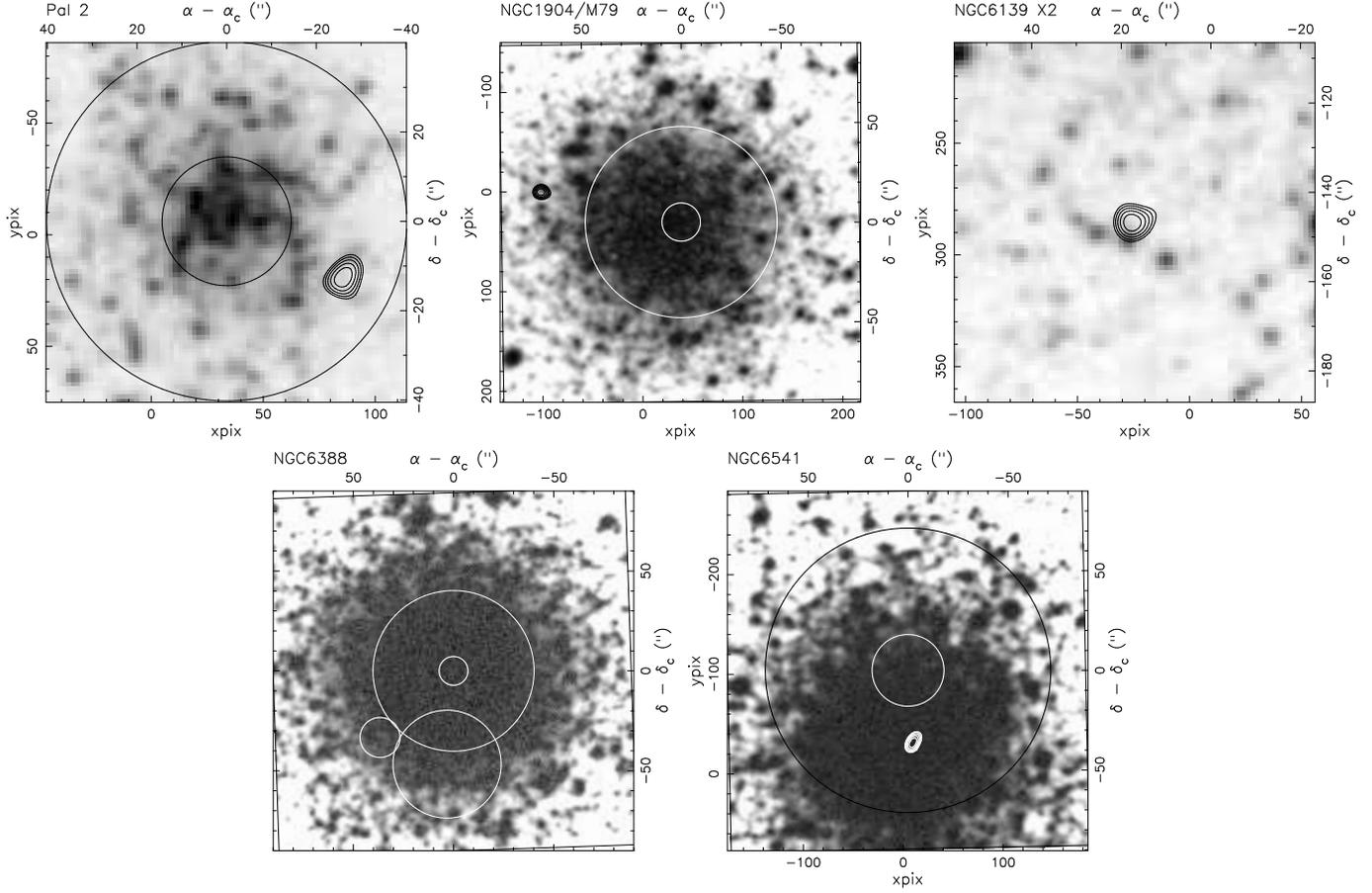

\centerline{
\parbox[b]{6.7cm}{\psfig{figure=figpal2.ps,width=6.cm,bbllx=120bp,bblly=60bp,%
bburx=494bp,bbury=435bp,clip=t}}
\parbox[b]{6.7cm}{\psfig{figure=fig1904.ps,width=6.cm,bbllx=120bp,bblly=60bp,%
bburx=494bp,bbury=435bp,clip=t}}
\parbox[b]{6.7cm}{\psfig{figure=fig6139.ps,width=6.cm,bbllx=120bp,bblly=60bp,%
bburx=494bp,bbury=435bp,clip=t}}
 }
\centerline{
\parbox[b]{6.7cm}{\psfig{figure=fig6388.ps,width=6.cm,bbllx=120bp,bblly=60bp,%
bburx=494bp,bbury=435bp,clip=t}}
\parbox[b]{6.7cm}{\psfig{figure=fig6541.ps,width=6.cm,bbllx=120bp,bblly=60bp,%
bburx=494bp,bbury=435bp,clip=t}}
 }
\caption{X-ray contours of ROSAT HRI observations, and X-ray positions
from PSPC observations, superposed on optical images from the Digitized 
Sky Survey. The contours were obtained after smoothing with a
2-d $\sigma$$=$3$''$ Gaussian. The inner circles indicate the 
core radii of the clusters, the outer circles the half-mass radii. 
The lower and left axes give pixel numbers for the ROSAT HRI detector,
the upper and right axes right ascension and declination with respect
to the cluster center. The conversions between pixel and celestial coordinates
are accurate to within $\sim5''$. Note the difference in scale of the images. 
a,b) the central areas of Pal\,2 and NGC\,1904, 
with X-ray sources outside the core radii at xpix,ypix of 86,19 and $-$100,0
respectively; c) the region near NGC\,6139\,X\,2;
d) NGC\,6388 with error circles for the position of 
the X-ray source as determined from the 1991 and 1992 data; 
e) NGC\,6541 with core and half-mass circles centered
on the cluster center as given by Djorgovski \&\ Meylan (1993), which 
apparently is offset from the actual center. The contours show the position
of the X-ray source.
\label{fover}}
\end{figure*}

The ROSAT HRI observation of Pal\,2 has been published by Rappaport et 
al.\ (1994), who detect a source at 6.5 core radii from the cluster center.
\nocite{rdlm94}
I find no other significant
sources in the observation, due to more stringent detection criteria than
applied by Rappaport et al.\ (1994).
The core radius of Pal\,2, as listed in Harris (1996), is larger than the
value used by Rappaport et al.\ (1994); the distance of the source to the
cluster center thus becomes 2 core radii.
As shown in Fig.\,\ref{fover}, the distance of the source to the cluster 
center puts it well inside the half-mass radius of the cluster.
I agree with the argument made by Rappaport et al.\ (1994) that
the source is probably associated with the cluster (see Sect.\,\ref{secprob}).

\subsection{NGC\,1904/M\,79}

One source is detected in the HRI observation of NGC\,1904, well outside the
half-mass radius (Fig.~\ref{fover}). I consider it likely that the source is 
associated with the cluster (see Sect.\,\ref{secprob}). The source was already
detected, with positional accuracy of 60$''$, by Hertz \& Grindlay
(1983) with the Einstein satellite.

\subsection{NGC\,3201}

Only one PSPC observation is available, which has been published by
Johnston et al.\ (1996).  \nocite{jvh96} 
Due to my more stringent significance limits,
I detect only three of the sources listed by Johnston et al., i.e.\ X\,5,
X\,6 and X\,7, and one new one.
No source is detected in the cluster.

\subsection{NGC\,4372}

Only one PSPC observation is available, which has been published by
Johnston et al.\ (1996). Due to my more stringent significance limits,
I detect fifteen of the sources listed by Johnston et al.;
due to use of the newer (improved) version of EXSAS, seven new sources
are detected.
None of the sources is in the cluster.

\subsection{NGC\,4590/M\,68}

No source is detected in the one available HRI observation.

\subsection{NGC\,5053}

Four sources are detected in the one available HRI observation; none
in the cluster core.

\subsection{NGC\,5139/$\omega$\,Cen}

The HRI observations of this cluster are described in Verbunt \&\ Johnston
(2000), who detect three sources in the core, the faintest of which may be a 
fore- or background source. The fluxes listed in Table\,\ref{tsum} are
those of source 9; for the HRI countrate I add the fluxes
of sources 9a and 9b.

\subsection{NGC\,5272/M\,3}
\label{s5272}

Two ROSAT HRI observations of this cluster were discussed by Hertz et 
al.\ (1993), who detected a relatively bright, very soft X-ray source 
in the first observation of Janunary 1992, which had faded below the detection
limit by July 1992. The source was also detected in the ROSAT All Sky
Survey (Verbunt et al.\ 1995), and previously with the Einstein satellite
(Hertz \&\ Grindlay 1983).
A complete overview of the ROSAT and ASCA observations
of this remarkable source is given by Dotani et al.\ (1999), who
discovered that the source has a $\sim$5\,keV bremsstrahlung spectrum
in its normal, low-luminosity state, and switches to a very soft 
($\sim$20\,eV for a blackbody fit) spectrum only when in the high-luminosity
state. \nocite{hgb93}\nocite{dag99}
I analyzed two HRI observations, viz. the longest one and the one in
which the source was brightest. Three sources are detected 
in addition to the cluster source in both observations, 
and all of these may be identified with quasars, as already noted by
Hertz et al. (1993) and Geffert (1998), who used their positions to 
determine the projection of the X-ray frame onto the optical sky. 
I repeat this determination, do it for the HRI images only,
and separately for the January 1992 and July 1995 observations.
In the 1992 observation, the bore sight correction has an uncertainty 
$\sigma_{\rm p}=1\farcs1$, and 
the position of the X-ray source on the frame has an uncertainty 
$\sigma_{\rm d}=0\farcs3$, leading to an overall uncertainty
in the position of the cluster source $\sigma_{\rm x}=1\farcs6$. 
In the longer 1995 observation in which the cluster source is much fainter,
$\sigma_{\rm p}=1\farcs0$ and $\sigma_{\rm d}=1\farcs1$, so that 
$\sigma_{\rm x}=1\farcs4$.
Thanks to the use of more accurate HRI positions, these accuracies are
better than those achieved before by Hertz et al.\ (1993) and Geffert (1998).
The positions listed in Table\,\ref{tsour} are those of the 1995 observation,
and take into account the shift which best brings the positions of B, C
and E into accordance with their optical identifications. 
The positions of the 1992 observation are compatible with these.
The source numbering is as
in Hertz et al.\ (1993), except for Source G, which is a new source,
not identical to source D listed by Hertz et al.\ (1993).
\nocite{hgp+92}\nocite{car76}\nocite{gef98}

\subsection{NGC\,5286}

As reported by Rappaport et al.\ (1994), the one HRI observation for this 
cluster detects no source in the cluster.

\subsection{NGC\,5466}

Three HRI observations of this cluster were obtained. No source
is detected in the core of the globular cluster in the
individual observations, or in the added image. 
An offset PSPC  observation also doesn't detect a cluster source;
the PSPC upper limit listed in Table\,\ref{tsum} is the countate of
a non-significant source detected at a distance of 1$'$ from the
cluster center.

\subsection{NGC\,5824}

No source is detected in either of the two HRI observations taken of
this cluster; the upper limit given in Table\,\ref{tsum} is from
the 1994 observation. This is slightly below the Einstein detection
at a level of $2\times10^{34}$\,\ergs in the 0.5-4.5\,keV range
(Hertz \& Grindlay 1983).

\subsection{NGC\,5904/M\,5}

In two short HRI observations of this cluster, no source is detected.
I therefore combine only the two longer HRI observations; ten sources are
detected in the resulting image, five (numbers 1-5 in Table\,\ref{tsour})
already given by Hakala et al.\ (1997), and five new ones (6-10).
One source may be identified with
a quasar, and one -- as noted before on the basis of the Aug 1994 observation
by Hakala et al.\ (1997) -- with the dwarf nova V\,101. (The position
for V\,101 listed in Table\,\ref{tsour} is that given by Evstigneeva
et al.\ 1995; Hakala et al.\ 1997 and
SIMBAD each give slightly different coordinates.)\nocite{esst95}
The offset between the X-ray coordinates and the optical coordinates
is determined from the two identified sources, and the result has been
applied to give the coordinates listed in Table\,\ref{tsour}. 
Sources X\,1 and X\,2 are also detected in the rather short PSPC observation,
which doesn't provide additional information.
The upper limits given in Table\,\ref{tsum} are for a position in the core.

\subsection{NGC\,5986}

The cluster is in the field of view of an HRI observation of the G6V star 
HIP\,77358. The star, but not the cluster, is detected.

\subsection{NGC\,6093/M\,80}

The ROSAT PSPC observation of this cluster shows one source in the center,
which may or not be the nova 1860 T Sco (Hakala et al.\ 1997).
The ROSAT image contains many sources, from which I retain in Table\,\ref{tsour}
only those with nominal positional accuracy better than 10$''$.
Three sources with good X-ray positions can be identified with stars. 
Two of these X-ray sources, those identified with the T Tau star VV\,Sco 
(an M1 star with $B = 12$, X\,1) and with HD\,146516 (a G1\,V star with 
$V=10.14$, X\,14) are listed as extended by the EXSAS software. 
I therefore determine the offset between X-ray and optical coordinates
from X\,7 only, identified with HD\,146457 (an  A5III/IV star with $V=8.46$).
The offset is applied to the X-ray positions, and the resulting positions
are listed in Table\,\ref{tsour}.
The uncertainties $\Delta$ in the source positions as listed in the
Table are the nominal uncertainties for the detector position only;
to these must be added in quadrature the random uncertainty
in positions for PSPC sources of 3$''$, which remains even after a
bore sight correction has been applied (Hasinger et al.\ 1998).
The uncertainty in the bore sight correction thus is $3\farcs5$ (assuming
a 1$''$ error in the optical position). 
The offset between the central source X\,10 and T Sco is $8\farcs7\pm6\farcs6$,
and the identification is possible.
The core radius of NGC\,6093 is about 9$''$, and the uncertainty of the
position of X\,10 is such that it could be identified with any source 
located in the cluster
core. I consider the identification of X\,10 with T Sco to be unproven.
\nocite{hbg+98}

\begin{figure*}
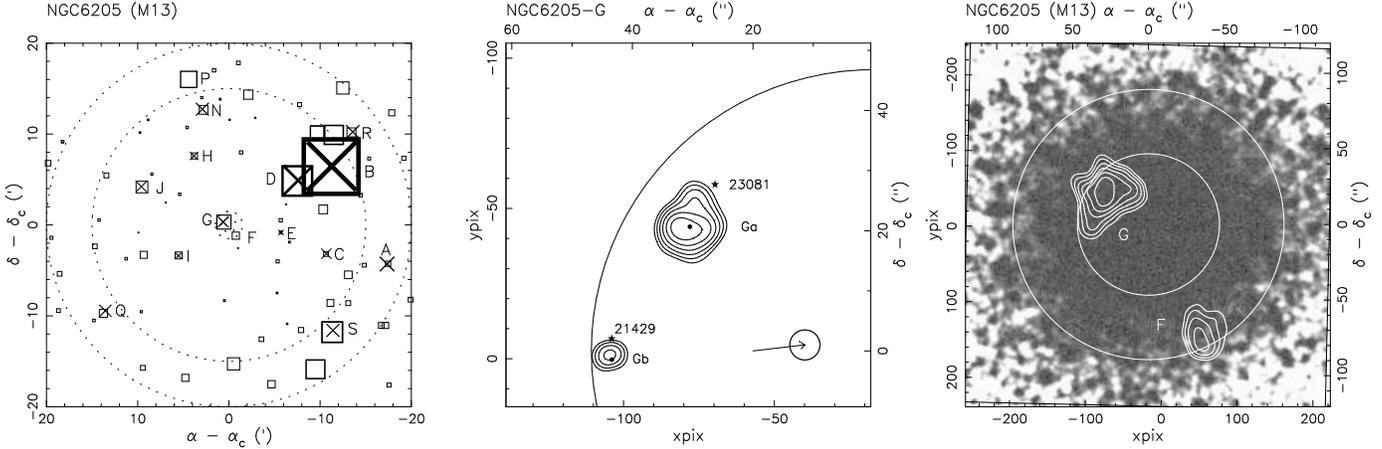

\centerline{
\parbox[b]{6.7cm}{\psfig{figure=fig6205a.ps,width=6.cm,bbllx=120bp,bblly=60bp,%
bburx=494bp,bbury=435bp,clip=t}}
\parbox[b]{6.7cm}{\psfig{figure=fig6205b.ps,width=6.cm,bbllx=120bp,bblly=60bp,%
bburx=494bp,bbury=435bp,clip=t}}
\parbox[b]{6.7cm}{\psfig{figure=fig6205c.ps,width=6.cm,bbllx=120bp,bblly=60bp,%
bburx=494bp,bbury=435bp,clip=t}}
 }
\caption{Results of ROSAT observations of NGC\,6205. Left: comparison between
the sources found with PSPC ($\Box$) and HRI ($\times$). The size of the
symbols is proportional to their countrate. The PSPC exposure, about twice
as long as the HRI exposure, detects all HRI sources and others in addition. 
The two small dotted 
circles show the core and half-mass radii of NGC\,6205; the two larger
circles have radii of 15$'$ and 20$'$ to help in comparing the effective
fields of view of PSPC and HRI (see Sect.\,\ref{secprob}). 
Middle: X-ray contours of source G from the HRI observation, obtained after
smoothing with a 2-d $\sigma$$=$2$''$ Gaussian. The positions of the two 
sources found with the multi-source algorithm are shown as $\bullet$;
two suggested ultraviolet counterparts (from Ferraro et al.\ 1997) are
indicated with $\star$. The circle segment delineates the core radius.
The arrow shows the size and direction of the displacement of the
X-ray contours with respect to the optical coordinate system that follows
if suggested identifications for X\,R and X\,T are accepted, the circle
indicating the uncertainty. Right: X-ray contours of sources G and F
from the PSPC observation,  obtained after smoothing with a 2-d 
$\sigma$$=$5$''$ Gaussian, superposed on the image of NGC\,6205 from the 
Digitized Sky Survey. The circles indicate the core and half-mass radii.
\label{fig6205}}
\end{figure*}

\subsection{NGC\,6121/M\,4}

Two ROSAT PSPC observations have been obtained of this cluster, one of
which was published by Johnston et al.\ (1996). Six sources were detected.
No source is detected in the cluster core in either PSPC observation.
Johnston et al.\ list 6 X-ray sources, one of which they identify with
the variable V972\,Sco $=$ HD\,147491 $=$  HIP\,80290, a G0\,V star with 
$V=9.7$. The Hipparcos catalogue lists the star as a suspected binary.
This source is also detected in each of the three ROSAT HRI observations 
obtained of NGC\,6121.
No source is detected in the cluster core in individual HRI observations;
I have combined the two longer observations, and detect a marginally
significant source in the core. The position of the X-ray frame is
fixed with help of X\,3/HD\,147491, and the uncertainty of the
core source is dominated by the uncertainty in its position on the
detector. To avoid confusion with the sources in Johnston et al.\ (1996)
I number the new sources as X\,7-10.

\subsection{NGC\,6139}

The short ROSAT HRI observation of this cluster has been published by
Rappaport et al.\ (1994); no source is detected in it. No source is
detected at the cluster position in two ROSAT PSPC observations pointed 
at white dwarf
WD\,1620$-$391; due to the large offset, the upper limit obtained from
these PSPC observations is not as good as that from the ROSAT All Sky
Survey.
I find two sources in the longer, hitherto unpublished HRI observation;
one of these is in the cluster core. The other one has no bright ($V<15$)
counterpart, but the Digitized Sky Survey provides several candidate
counterparts, as shown in Fig\,\ref{fover}. 
Spectroscopic confirmation of any of these would provide
improved accuracy in the position of the cluster source, by determining
the bore sight correction.

\subsection{NGC\,6205/M\,13}

A ROSAT HRI observation and two PSPC observations of this cluster were
analyzed by Fox et al.\ (1996). \nocite{flm+96} 
The new analysis of the HRI observation gives essentially the same sources,
at slightly different positions due to the new value for the pixel scale.
One new source, R, is detected. 
Sources F, Q and S are not significant in the HRI according to the ML$>$13
criterium, but are listed because they are significantly detected in the
PSPC. F, within the half-mass radius, and S are new sources.
A comparison between
the sources detected with the two detectors is shown in Fig.\,\ref{fig6205}.
Outside the area analyzed by Fox et al.\ (1996),
one source, listed as T in Table\,\ref{tsour}, has a relatively accurate
position.

None of the sources listed in Table\,\ref{tsour} has an optical counterpart
in SIMBAD; inspection of the Digitized Sky Survey images gives several 
candidate identifications, including a star of $B$$=$13.1 with X\,R and an 
extended source (galaxy?) of $R=14.8$ with
X\,T. All other X-ray sources have counterparts with $B,R\gtap15$.
Since X\,T is outside the HRI field of view, I determine the bore sight
corrections as follows. 
Comparison of the positions of the sources common to PSPC and HRI shows that a 
correction of $-0\fs02,+2\farcs2$ added to the PSPC positions brings them
in line with the HRI positions. If the identifications for X\,R and X\,T are 
correct, a correction of $-0\fs74,+3\farcs3$ added to
the PSPC coordinates brings them to the optical coordinates.
Thus a correction of $-0\fs72,+1\farcs0$ must be added to the HRI
coordinates listed in Table\,\ref{tsour} to bring them to J\,2000.
This correction, with its error circle of radius $2\farcs5$, is indicated
in Figure\,\ref{fig6205}.
The positions given in Table\,\ref{tsour} are uncorrected HRI positions, 
because I consider the suggested optical identifications too uncertain;
the positions of X\,F and X\,T, found in the PSPC only, have been corrected
to the HRI frame.

The multiple-source algorithm detects two significant sources in the cluster
core, labelled as Ga and Gb in Table\,\ref{tsour}. Gb is the 3.5-$\sigma$
source noted by Fox et al.\ (1996), labelled G-SE by Ferraro et al.\ (1997).
\nocite{fpf+97b} Given the uncercainty of the bore sight
correction, the position of star 23081 is compatible with Ga, and the
position of star 21429 with Gb. However, the distance between the positions
of stars 23081 and 21429 is not compatible with the distance between Ga and Gb.
Thus, if we accept the identification of star 21429 with Gb, star 23081 cannot
be identified with Gb, and vice versa.
The optical identifications of X\,R and X\,T  result in the bore sight 
correction indicated with an arrow in Fig.\,\ref{fig6205}, which would exclude 
the identifications of star 23081 with Ga and of star 21429 with Gb.

\begin{figure*}
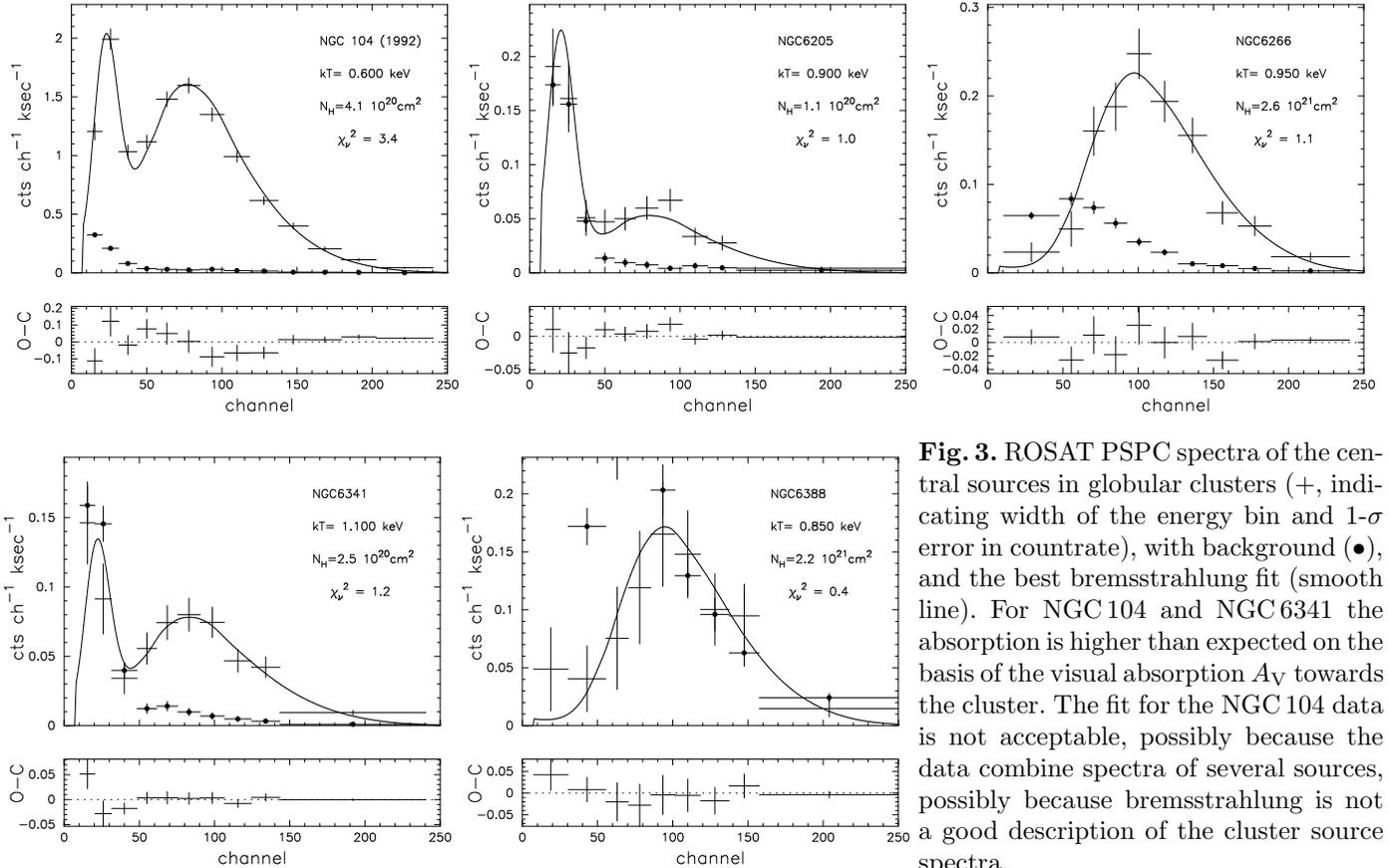

\centerline{
\parbox[b]{6.7cm}{\psfig{figure=fsp104.ps,width=6.cm,bbllx=40bp,bblly=270bp,%
bburx=544bp,bbury=741bp,clip=t}}
\parbox[b]{6.7cm}{\psfig{figure=fsp6205.ps,width=6.cm,bbllx=40bp,bblly=270bp,%
bburx=544bp,bbury=741bp,clip=t}}
\parbox[b]{6.7cm}{\psfig{figure=fsp6266.ps,width=6.cm,bbllx=40bp,bblly=270bp,%
bburx=544bp,bbury=741bp,clip=t}}
}
\centerline{
\parbox[b]{6.7cm}{\psfig{figure=fsp6341.ps,width=6.cm,bbllx=40bp,bblly=270bp,%
bburx=544bp,bbury=741bp,clip=t}}
\parbox[b]{6.7cm}{\psfig{figure=fsp6388.ps,width=6.cm,bbllx=40bp,bblly=270bp,%
bburx=544bp,bbury=741bp,clip=t}}
\parbox[b]{6.2cm}{
\caption{ROSAT PSPC spectra of the central sources in globular clusters
($+$, indicating width of the energy bin and 1-$\sigma$ error in countrate), 
with background ($\bullet$), and the best bremsstrahlung fit 
(smooth line). For NGC\,104 and NGC\,6341 the absorption is higher than
expected on the basis of the visual absorption $A_{\rm V}$ towards the
cluster. The fit for the NGC\,104 data is not acceptable, possibly because
the data combine spectra of several sources, possibly because bremsstrahlung is
not a good description of the cluster source spectra.
\label{fspec}}}
}
\end{figure*}

The PSPC spectrum of source G, which includes counts of both Ga and Gb, is
adequately fitted by a bremsstrahlung spectrum, as shown in Fig.\,\ref{fspec}.
The background spectrum was determined from a circle located north-west 
of the source, to avoid source F.

\subsection{NGC\,6254/M\,10}

The ROSAT PSPC observation of this cluster has been published by Johnston et 
al.\ (1996), and doesn't show a source in the cluster core.
I detect two sources in addition to those found by Johnston et al.\ (1996).

\subsection{NGC\,6266/M\,62}

Two PSPC observations of the low-mass X-ray binary, burster and transient,
MXB\,1659$-$29 contain this globular cluster in the field of view.
The transient is not detected in either observation,
but the globular cluster is in both! 
(The upper limit to the countrate of MXB\,1659$-$29
in channels 50-240 is 0.6\,\ctks\ in 1991, 0.9\,\ctks\ in 1992; for
$\nh\simeq2\times10^{22}$\,\cmsq -- as determined from the X-ray spectrum
during the active stage by Wachter et al.\ 2000 --
and an assumed 0.3\,keV blackbody spectrum -- typical for a quiescent
transient -- the 1991 limit corresponds to an upper limit to the luminosity 
in the 0.5-2.5 keV band of about $1.4\times10^{33}$\,\ergs at 10\,kpc.)
The countrate and spectrum of the cluster source is not significantly 
different between the 1991 and 1992
observations. I therefore analyse the spectrum of the combined image.
As shown in Fig.\,\ref{fspec}, the spectrum is soft, and affected by
strong absorption. Fixing the absorption column at
$\nh=2.6\times10^{21}$\,\cmsq, as indicated by the visual absorption
of the cluster,
I find that a bremsstrahlung spectrum of 0.95$\pm$0.17\,keV describes the
spectrum; the observed countrate corresponds to a
luminosity in the 0.5-2.5\,keV range of $\lx\simeq1.6\times10^{33}$\,\ergs
at the distance of 5\,kpc of the cluster. Thus spectrum and luminosity
of the source are similar to what is observed in soft X-ray transients
with a neutron star in quiescence (e.g.\ Aql X-1, Verbunt et al.\ 1994).
\nocite{wsb00}\nocite{vbj+94}

Two sources in the field of view and with well determined positions of the 
ROSAT PSPC have an counterpart at other wavelengths: X\,8 with the
radio source PKS\,1657$-$298 and X\,12 with the star TYC\,7360\,394\,1 
(CD-29\,13096),
a binary with $V=10.1$. The positions of the X-ray sources
coincide within 3$''$ and 1$''$ with the Parkes radio source and Tycho
Catalogue object, respectively. Since both X\,12 and the cluster source
X\,14 are indicated to be extended sources by the EXSAS software
in the individual 1991 and 1992 images, the actual accuracy of the 
position may be worse than the positional coincidences suggest; and I have
not applied a bore sight correction based on these identifications.
Confirmation that both sources are extended would be interesting, both 
for the binary and for the cluster source.
\nocite{cw89}\nocite{wwt+91}\nocite{hbb+97}\nocite{plk+97}

\subsection{NGC\,6273/M\,19}

The ROSAT HRI observation of this cluster has been published by
Rappaport et al.\ (1994); no source is detected in it.

\subsection{NGC\,6293} 

Two PSPC observations of the cataclysmic variable V\,2051 Oph have this
cluster near the edge of the detector; a bright source about 10$'$ from 
the cluster prevents the determination of a useful upper limit.
This bright source has a countrate of about 2.5\,\cts, and probably is 
identical to 1RXS\,170930.2$-$263927. The pointed observations, as well
as the Survey, indicate that the source is extended. Its spectrum is absorbed,
all its counts detected in channels 50-240, i.e.\ at energies $\gtap0.5$\,keV.
SIMBAD gives IW\,Oph, also known as (Harvard Variable) HV\,4409 as a
possible counterpart, whose nominal position -- with 1$'$ uncertainty --
is compatible with the X-ray source; however, inspection of the finding chart 
in Swope (1932) shows that HV\,4409 is too far from the X-ray position
to be an acceptable counterpart for the X-ray source.

\subsection{NGC\,6304}

I confirm the number of counts detected in this cluster by Rappaport et 
al.\ (1994); but consider them as representing an upper limit rather than 
a detected source, since no significant source is detected in the cluster
by EXSAS, or by my own multi-source software. 
The upper limit is given in Table\,\ref{tsum}.

\subsection{NGC\,6316}

No source is detected in the ROSAT HRI observation of this cluster,
as pointed out by Rappaport et al.\ (1994).

\subsection{NGC\,6341/M\,92}

A short PSPC observation of this cluster has been published by Johnston et 
al.\ (1994), and longer HRI and PSPC observations by Fox et al.\ (1996).
I analyze the two HRI observations separately, determine the offset between
them, and then analyze the co-added image, to find positions which differ
markedly from those listed in the earlier papers. \nocite{jvh94}

In agreement with Geffert (1998) I identify two X-ray sources in the HRI field
of view with optical objects, viz.\ X\,E with the W UMa variable V798\,Her
(V\,14 of Hachenberg 1939),
whose accurate position is given by Tucholke et al.\ (1996), and X\,10 with
TYC\,3081\,510\,1 from the Tycho Catalogue which also has an accurate position.
The binary period of V798\,Her is 0.346\,d, and from this I estimate
$M_{\rm V}\simeq4.5$; with the observed magnitude $V=14.5$ a distance of order 
1\,kpc follows, and the observed  countrate then corresponds to a luminosity 
between 0.1-2.4\,keV 
$\lx\simeq 10^{30}$\,\ergs, quite reasonable for a contact binary
(see, e.g.\ McGale et al.\ 1996). X\,E is detected only in the longest
PSPC and in the longest HRI observation; compatible with a constant flux.
Geffert (1998) suggests on the basis of its colour that TYC\,3081\,510\,1
is a G star; it would then have an X-ray luminosity high
but not unreasonable for a G main sequence star (see Table\,\ref{tstar}).
Its X-ray flux increases by about 40\%\ between April 1994 
(HRI countrate 25.6$\pm$1.3\,\ctks) and April 1995 
(HRI countrate 43.2$\pm$4.1\,\ctks); its countrates in the long PSPC
observation of 1992 are 54.8 $\pm$1.1 and 106.7$\pm$1.6\,\ctks\ in channels
50-240 and 11-240, respectively; the large countrate in the lower channels
indicates that interstellar absorption is low, as expected for a stellar
X-ray spectrum at the small distance of the optical counterpart.
Thus, the X-ray countrates are compatible with the suggested identifications
for X\,E and for X\,10.

From these two objects I determine the offset -- given in Table\,\ref{tobse} --
between the X-ray coordinates and the optical coordinates, and apply this
to the X-ray coordinates to obtain the positions given in Table\,\ref{tsour}.
The source numbering in the Table is after Johnston et al.\ (1994) and 
Fox et al.\ (1996).
The offset determination is dominated by X\,10, because of its
accurate position on the X-ray detector (error $\Delta$ in Table\,\ref{tsour}).
The estimated errors in the offset are $0\farcs6$ in right ascension and
in declination; combining these with the error in the position on the
detector of source X\,8, I estimate that the position of X\,8 is accurate 
to $1\farcs5$, comparable to the accuracy obtained by Geffert (1998).
Note, however, that for non-identified sources the HRI positions given in
Table\,\ref{tsour} are more accurate than those given by Geffert (1998), which
were based on PSPC positions.
\nocite{gef98}\nocite{flm+96}\nocite{tsb96}\nocite{mph96}\nocite{hsv98}

The PSPC countrate of X\,8 is 3.0$\pm$0.7\,\ctks\ (channels 50-240) in
the short PSPC observation, compatible with expectation on the basis
of its HRI countrate; during the long PSPC observation it is
substantially higher, at 6.5$\pm$0.4. The source also has an appreciable
countrate in the softer channels, and a bremsstrahlung
fit is shown in Fig.\,\ref{fspec}. If we set $\nh=10^{20}$\,\cmsq, as
estimated from the visual absorption of the cluster, 
we cannot obtain a good fit; the
absorption to X\,8 requires $\nh\simeq2.5\times10^{20}$\,\cmsq. 
The spectrum is
soft, with a bremsstrahlung temperature $kT=1.1\pm0.2$\,keV.

It should be noted that the identification of X\,8 (called source C 
in Fox et al.\ 1996) with an ultraviolet object UV\,8203
by Ferraro et al.\ (2000) is based on the old, inaccurate positions of Fox 
et al.\ (1996), which are incompatible both with the improved positions
of the PSPC sources by Geffert (1998) and of the HRI sources in this paper.
These positions indicate that the proposed identification is not acceptable.
\nocite{fpr+00}

\subsection{NGC\,6352}

In re-analyzing the PSPC observation of this cluster I confirm the absence 
of a source in the cluster core, and note that source X\,4 of Johnston
et al.\ (1996) is within the half-mass radius of the cluster.
I also confirm (with X-ray positions improved by the new analysis software)
the identity of X\,9 with HD\,157522 (HIP\,85326) and X\,10 with HD\,157555
(HIP\,85342). 
HD\,157522 is a K1\,V star at 47\,pc.
HD\,157555 is a F6-7V star at 46.5\,pc.
The X-ray luminosities of both stars are $\sim 10^{29}$\ergs, as expected
for such stars (see Table\,\ref{tstar} and Fig.\,\ref{frass}).

These two identifications are used to determine the bore
sight correction, as listed in Table\,\ref{tobse}, leading to the
positions given in Table\,\ref{tsour}. I only list sources with positional
accuracy better than 10$''$; numbers up to 17 are those of Johnston et al.\ 
(1996), source X\,17 is new.
The accuracy of the bore sight correction is $2''$. 
At its low countrate and large ($90''$) distance from the cluster
center, X\,4 may well be a fore- or background object (Sect.\,\ref{secprob}); 
its positional accuracy is dominated by its $\Delta$, as given in 
Table\,\ref{tsour}.

\subsection{NGC\,6366}

The one PSPC observation of this cluster was published by Johnston et al.\ 
(1996), who detect 6 sources in channels 11-240; I confirm their results,
and detect two new sources by limiting the analysis to channels 50-140.
These sources are listed as no.s 7,8 in Table\,\ref{tsour}. 
A faint source is detected in the core of the cluster, with a rather
uncertain position.

The brightest source in the field of view is the F3V star HD\,157950
(HIP\,85365), already detected as an Einstein source and in the ROSAT
All Sky Survey.  Its distance is 30\,pc, and it is a binary with a
period of 26.2765\,d, eccentricity $e=0.49$, and radial velocity
semi-amplitude $K_1=47.5$\,km/s (orbit by Parker 1915, listed as
good but not definitive by Batten et al.\ 1989). \nocite{bfm89}\nocite{par15}
This source was used to determine the bore sight correction given in
Table\,\ref{tobse}. The uncertainty in the bore sight correction is about
3$''$.

\subsection{HP\,1/ESO 455-11}

A short HRI observation of this highly reddened, collapsed cluster
shows only one source, not related to the cluster.

\subsection{NGC\,6380}

A short HRI observation, reported by Rappaport et al.\ (1994), shows no
source in the cluster core.

\subsection{NGC\,6388}

This cluster is in the field of view of two PSPC observations, one pointed
at the bright low-mass X-ray binary 4U1735$-$44, and a rather shorter one
pointed at Gliese\,682.
A source near the globular cluster is detected in both observations.
I have analysed part of the field-of-view of the 1991 observation, 
and determine the bore-sight correction on the basis of identifications of 
X\,1 with TYC\,7896\,3885\,1 and of X\,2 with HIP\,86356.
Applying this correction I obtain the positions listed in Table\,\ref{tsour}.
The resulting position of X\,5 is just outside the half-mass radius of
NGC\,6388 (see Fig.\,\ref{fover}), and has a positional uncertainty of 10$''$.
Thus the source may or may not be related to the cluster.

The 1992 observation contains three sources with accurate detector positions:
X\,1 and two sources that may be identified with TYC\,7896\,3812\,1 and 
TYC\,7896\,2299\,1 respectively. With the bore sight determined from these
three sources I obtain a position for X\,5 which is compatible with that
from the 1991 observation, but with a much larger error of about 30$''$
(Table\,\ref{tsour}).

If X\,5 belongs to NGC\,6388, its soft photons should be absorbed. 
The countrate over the full PSPC range of channels 11-140 indeed is not
higher than that in the hard range channels 50-240 only (listed in 
Table\,\ref{tsour}), as expected for a cluster member.
X\,5 is variable, being about 50\%\ brighter in 1992 than in 1991. When fitted
with a brems strahlung spectrum, and fixing $\nh$ at the value suggested by 
$A_{\rm V}$ of the cluster,
i.e. $\nh=2.2\times 10^{21}$\,\cmsq, the temperature is found
to be $0.85^{+0.5}_{-0.2}$\,keV, and the source luminosity between 
0.5-2.5\,keV is $6\times10^{33}$\,\ergs in 1991.

In contrast, X\,1 and X\,2 have countrates in channels 11-240
higher than their countrates in channels 50-240, as expected for foreground 
stars. 
TYC\,7896\,3812\,1 may
be a G5\,V star at about 100\,pc, with an X-ray luminosity at the
highest end of the distribution among nearby G5\,V stars (Table\,\ref{tstar}).
HD\,159808 (HIP\,86356) is a binary with spectral type F5\,V, which with 
$V$$=$8.12 puts its at about 100\,pc (spectroscopic parallax), compatible
with the astrometric parallax which converts to a distance of 150$\pm$40\,pc.
Its X-ray luminosity is at the high end of the 
distribution of X-ray luminosities for nearby F5\,V stars (Table\,\ref{tstar}).
TYC\,7896\,3812\,1 could be a 
K1\,V star at 75\,pc, with an X-ray luminosity of $1.7\times10^{30}$\,\ergs.
The X-ray counterpart of TYC\,7896\,2299\,1 has no photons at soft energies, 
indicating that some interstellar absorption is present.
Its colour indicates spectral type earlier than K5\,V at a distance larger 
than 36\,pc (Table\,\ref{tstar}).
The X-ray luminosities derived for all proposed optical counterparts
are as expected for such stars.
Foreground stars X\,1 and X\,2 are both fainter in 1992 than in 1991, by
30\%\ and 40\%, respectively.

\subsection{Djorg\,1}

This cluster is in the field of view of an HRI observation of 
RX\,J1748.9$-$3254, but is not detected.

\subsection{Terzan\,6}

The transient in this cluster was detected in outburst during the ROSAT All
Sky Survey (Predehl et al.\ 1991, Verbunt et al.\ 1995); but not detected in
an HRI pointing obtained in March 1992 (Rappaport et al.\ 1994).
The latter observation thus provides an upper limit for dim sources in this 
cluster.
\nocite{phv91}

\subsection{NGC\,6453}

The HRI observation made of this cluster shows no source near the cluster,
as already reported by Rappaport et al.\ (1994).
I have analysed a PSPC observation of the open cluster NGC\,6475/M\,7
(published by James \&\ Jeffries 1997) and from
it find an upper limit at the position of the cluster center, which is equal
to the detected rate for a source located 2$'$ away from it.
\nocite{jj97}

\subsection{NGC\,6496}

I confirm the absence of a cluster source in the PSPC observation
reported by Johnston et al.\ (1996).
Note that the upper limit in the hard band, as given in Table\,\ref{tsum},
is lower than that for the total band given by Johnston et al.\ (1996).

\subsection{NGC\,6522 and NGC\,6528}

A marginally significant (ML$=$12.5) source near NGC\,6522 is detected 
at $\alpha=18^h03^m35\fs92$, $\delta=-30^{\circ}02'07\farcs8$ in a 
PSPC observation of Baade's window, well outside the core radius, but 
inside the half-mass radius. The uncertainty in the source position
is about 10$''$. The source is not detected in an analysis of all
energy channels (11-240), showing up only in the hard band (channels 50-240).
The upper limit given in Table\,\ref{tsum} for a source
in the core is the countrate of this source.
NGC\,6528 is in the field of view of the same PSPC observation,
but it is not detected. Its much improved upper limit is listed in 
Table\,\ref{tsum}.

\subsection{NGC\,6540/Djorg\,3}

This cluster is in the field of view of a PSPC observation. A source
is detected at  $\alpha=18^h06^m11\fs97$, $\delta=-27^{\circ}45'55\farcs6$,
45$''$ from the cluster center, but it has a significant countrate in
the soft channels, and thus must be a foreground object, because
the cluster is highly absorbed. The upper limit given in Table\,\ref{tsum}
is the countrate of this source.

\subsection{NGC\,6541}

Two sources are detected in the HRI observation of this cluster, one
near the cluster center. The latter source is probably identical to the
source detected near NGC\,6541 with the Einstein satellite (Hertz \&\ Grindlay
1983).
X\,1 may be identified with TYC\,7911\,112\,1, possibly a K2\,V star at
a distance of about 65\,pc (Table\,\ref{tstar}). The position of this source 
is used to determine the bore sight correction. The coordinates of the cluster
center as given by Djorgovski \&\ Meylan (1993) suggest that X\,2 is outside 
the cluster core, but overlaying the X-ray image on the Digitized Sky Survey
shows that it is, in fact, near the cluster center, as illustrated in
Fig.\,\ref{fover}. \nocite{dm93}

\subsection{NGC\,6544}

No source near the cluster is detected in the PSPC observation of this
cluster, as already reported by Johnston et al.\ (1994).

\subsection{NGC\,6553}

The PSPC observation of RXJ\,1806.8$-$26.06
which has this cluster in the field of view shows no 
source at the position of the cluster.

\subsection{Ter\,11}

The cluster is in the field of view of a PSPC observation of G8.3$-$1.7, 
from which the upper limit given in Table\,\ref{tsum} is found.

\subsection{NGC\,6626/M\,28}

\begin{figure*}
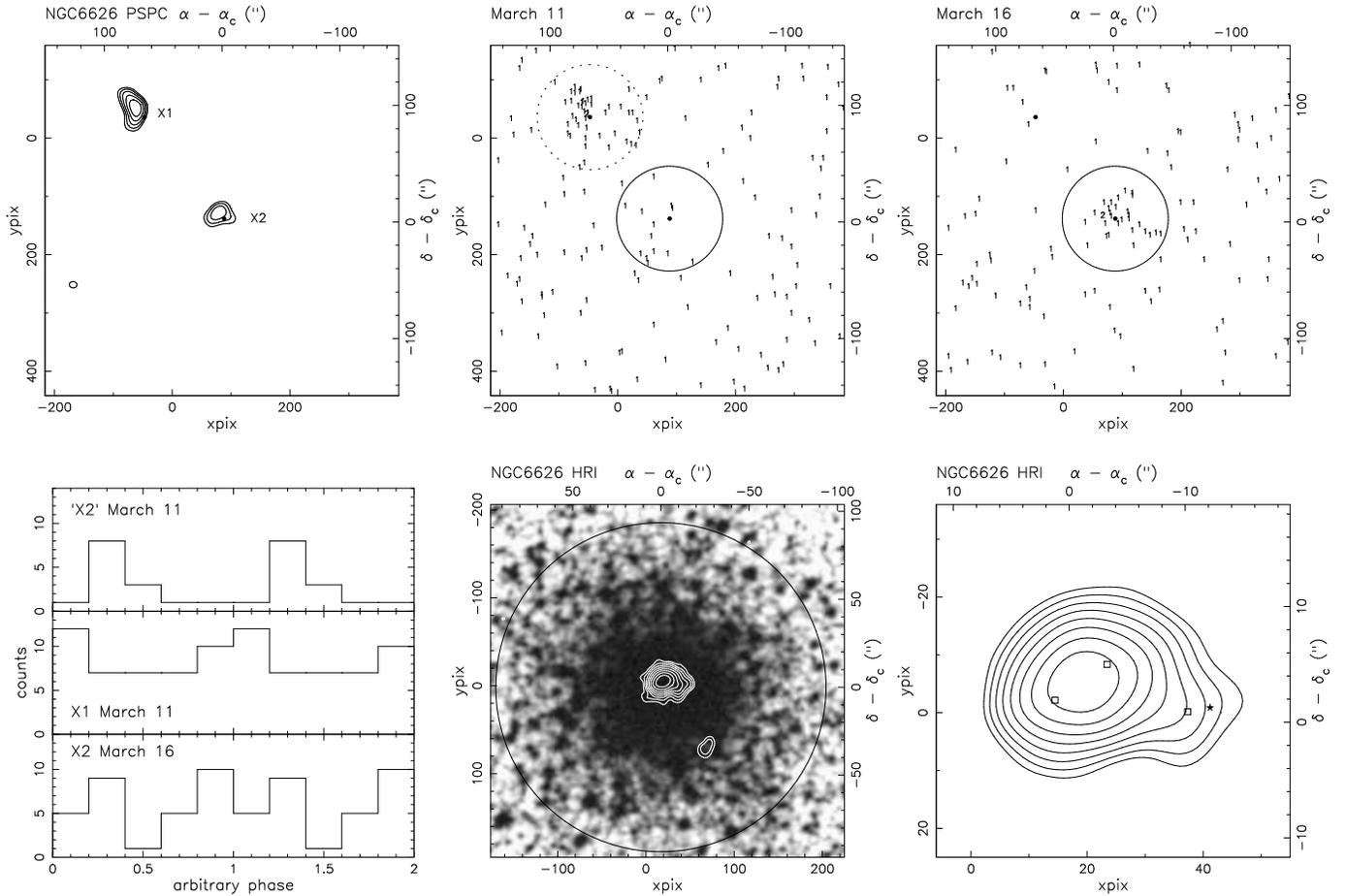

\centerline{
\parbox[b]{6.7cm}{\psfig{figure=fpulsa.ps,width=6.cm,bbllx=120bp,bblly=60bp,%
bburx=494bp,bbury=445bp,clip=t}}
\parbox[b]{6.7cm}{\psfig{figure=fpulsb.ps,width=6.cm,bbllx=120bp,bblly=60bp,%
bburx=494bp,bbury=445bp,clip=t}}
\parbox[b]{6.7cm}{\psfig{figure=fpulsc.ps,width=6.cm,bbllx=120bp,bblly=60bp,%
bburx=494bp,bbury=445bp,clip=t}}
}
\centerline{
\parbox[b]{6.7cm}{\psfig{figure=m28pulse.ps,width=6.cm,bbllx=40bp,bblly=60bp,%
bburx=414bp,bbury=455bp,clip=t}}
\parbox[b]{6.7cm}{\psfig{figure=n6626.ps,width=6.cm,bbllx=120bp,bblly=60bp,%
bburx=494bp,bbury=435bp,clip=t}}
\parbox[b]{6.7cm}{\psfig{figure=f6626h.ps,width=6.cm,bbllx=120bp,bblly=60bp,%
bburx=494bp,bbury=435bp,clip=t}}
}
\caption{Above: PSPC images of the inner area of NGC\,6626. From left to 
right: 
a) total image, smoothed with 2-d Gaussian with width 6$''$, showing the two 
sources X\,1 and X\,2 from Johnston et al.\ (1994) near the cluster center, 
b) locations on the detector of photons from March 11,
c) locations on the detector of photons from March 16.
Only X\,1 is detected in the March 11 data, 
only X\,2 in the March 16 data, and I conclude that X\,1 is the cluster source
shifted due to (slightly) incorrect satellite attitude solution for March 11.
$\bullet$ gives the positions of X\,1 and X\,2, the full circles indicate
an area around X\,2 with radius 45$''$ from which Danner et al.\ (1994)
extracted 'pulsar' photons, whereas the dotted circle indicates the area from
which photons should be extracted for March 11.
Below, from left to right:
d) photons from the circles shown in b) and c) folded on the pulse period of
PSR\,B1821$-1$24
e) ROSAT HRI image, smoothed with a 2-d Gaussian with width 2$''$ superposed
on an image from the Digitized Sky Survey. The circle gives the half-mass 
radius, X\,7 is well within this radius
f) X-ray contours of the core source in NGC\,6626, smoothed with
a 2-d Gaussian with width 2$''$, of the co-added HRI image, with exposure
of 80\,ksec. The positions of the sources found with the multiple-source
fit are indicated with $\Box$, from left to right X\,2a, 2b and 2c; the
position of PSR\,B1821$-$24 is shown with $\star$. The conversion of detector
coordinates (lower and left axes) to J2000 coordinates (upper and right axes)
is accurate to about 2$''$, thus the position of X\,2c is compatible
with that of the pulsar.
\label{fpuls}}
\end{figure*}

The PSPC observation of this cluster has been published by Johnston et 
al.\ (1994), who detect four sources, two -- X\,1 and X\,2 -- 
near the cluster core. I will show that X\,1 and X\,3 are spurious.
Danner et al.\ (1994) claimed to detect the pulse period of PSR\,B1821$-$24 
in X\,2. I will show that this pulse detection is spurious.
The actual detection of pulsed emission from NGC\,6626 must be credited
to Saito et al.\ (1997), who used ASCA data.
A long ROSAT HRI observation was obtained by Danner et al.\ (1997),
who show that X\,2 is a multiple source, and that the pulsed source
is off center. Their preliminary analysis assigns one third of the
flux of X\,2 to the pulsar; a detailed analysis below shows that
the actual pulsar flux is lower.
\nocite{dkt94}\nocite{dksk97}\nocite{skk+97}

\subsubsection{NGC\,6626 PSPC observations}

The PSPC data were obtained on March 11 (effective exposure of 1911\,s)
and on March 16 (1423\,s). Analyzing the data of these two days
separately, I find that X\,1 is present on March 11 but not on March 16;
and that X\,2 is present on March 16 but not on March 11, as illustrated
in Fig.\,\ref{fpuls}.
I conclude from this that the two halves of the total exposure
are offset with respect to one another, and that X\,1 and X\,2 are in fact
the same source. This is supported by the fact that their countrates
are the same: 21.1$\pm$3.6\,\ctks\ for X\,1 on March 11 and 
18.4$\pm$3.8\,\ctks\ for X\,2 on March 16. (There is no other source
in the field of view with which the shift can be confirmed independently.)
The 13 counts analysed by Danner et al.\ (1994) for March 11
are background photons, not related to X\,2. This statement is further
supported by the facts, illustrated in Fig.\,\ref{fpuls}, that a)
the distribution of the photons of X\,2 on March 11 
doesn't follow the point spread function, and
b) the background map for March 11 predicts 9 background photons in
an extraction circle with a radius of 45$''$, as used by Danner et al.\ (1994).
Even if we assume with Danner et al.\ (1994) that there were no offset,
and with less than a third of the counts of X\,2 being due to the pulsar, it
follows that only 1 or 2 of the 13 photons would belong to the pulsar.
Clearly the pulse detection in these data by Danner et al. (1994) is 
spurious.
Fig.\,\ref{fpuls} shows the 13 background photons folded on the
pulsar epheremis and confirms the result by Danner et al.\ (1994) that one
of five bins used contains 8 photons. 
(The figure in fact shows 14 photons from a slightly larger extraction radius;
this doesn't affect the discussion below.)
With 13 photons distributed randomly 
over five bins, the binomial probability of finding a bin with 8 or
more photons is 0.006, a factor 20 higher than the probability obtained
by Danner et al.\ (1994) 'through a Monte Carlo simulation'.

I conclude that source X\,1 from Johnston et al.\ (1994) is an image of
X\,2, and does not exist as a separate source. Source X\,3 is
not detected in the separate observations of March 11 and March 16, and
I consider it spurious as well.
The correct countrate for X\,2 is found from the sum of the counts of X\,1 and
X\,2 as given by the standard analysis. 
This corrected rate is given in Table\,\ref{tsum}.

\subsubsection{NGC\,6626 HRI observations}

Four HRI observations of this cluster have been obtained. The 1995 observation
has been published by Danner et al.\ (1997), who show that the
central source X\,2 is multiple.
In analyzing this observation I detect 3 sources in addition
to the central source, but these sources are not detected in the remaining
three, shorter HRI observations. To add the images, I therefore used the
position of the central, extended source; this implies the assumption that the 
X-ray center of this source doesn't vary.
The shifts applied to align the three shorter images with the longer one
are given in Table\,\ref{tobse}.
The co-added image contains 9 significant sources, including the central
source X\,2 and the three additional sources already detected in the first HRI
observation, i.e.\ X\,5, X\,9 and X\,12. These three sources have higher
significance in the co-added image than in the first observation only, and
improved positional accuracy, which indicates that the addition of
the images is correct.

X\,9  can be identified with HD\,315622 (TYC\,6848\,3536\,1) and this star 
is used to correct the X-ray coordinates, giving the positions listed
in Table\,\ref{tsour}. 
The right ascension and declination of X\,9 determined for only 1995 data
and for the total image differ by $2\farcs0$ and $0\farcs4$, respectively, 
from which I conclude
that the coordinate frame of the total image is accurate to about 2$''$. 
No significant variability of the flux of X\,9 is indicated by
the non-detection of this source in the three shorter HRI observations,
which may be explained with the shorter exposure times.

The central source X\,2 is significantly variable between the first HRI
observation, in which the countrate is 11.8$\pm$ 0.6\,\ctks\ and the remaining 
three observations, in which the countrates are 7.7$\pm$1.4, 6.5$\pm$0.5 and
7.7$\pm$1.0\,\ctks, respectively.
X\,7 is a source outside the core but within the half-mass radius; it is
variable, with a countrate of 0.3$\pm$0.1\ctks\ in 1995 and 
0.8$\pm$0.2\ctks\ in 1996 Sep.
X\,2 and X\,7 are shown superposed on the optical image in Fig.\,\ref{fpuls}.

The multiple-source fit to the central 50$''\times$50$''$ gives a total of 
three significant sources, with positions and countrates as listed in 
Table\,\ref{tsour}. \nocite{tml93}
An image is given in Fig.\,\ref{fpuls}.
Multiple-source fits to only the 1995 data and to only the 1996 Sep data give 
source positions compatible with those for the fit to the co-added total data.
X\,2a is significantly variable, with countrate 6.2$\pm$0.6\,\ctks\ in 1995 and
1.6$\pm$0.5\,\ctks\ in 1996 Sep, respectively, in agreement with the observed 
variation in the total countrate of X\,2.
The added countrates of sources X\,2a-c are compatible with the single-source
countrates determined for X\,2 by the standard analysis.

The position of X\,2c coincides within the error with PSR\,B1821-24, taking
into account the overall uncertainty of 2$''$ in the bore sight correction.
The multiple-source fits indicate that the pulsar has a countrate of
1.5$\pm$0.2\,\ctks, contributing about 17\%\ of the photons of X\,2.

\subsection{NGC\,6638}

Two PSPC observations have this cluster in the field of view; the best
upper limit, given in Table\,\ref{tsum}, is from the longer observation.

\subsection{NGC\,6642}

Two PSPC observations have this cluster in the field of view. No source
is detected in the cluster. The upper limit from the observation pointed
at the cluster, reported by Johnston et al.\ (1994), is not improved by the
longer observation pointed at an offset.

\subsection{NGC\,6656}
\label{s6656}

A source in the core of this cluster has been found with the PSPC,
as reported by Johnston et al.\ (1994). 
Three sources are detected in the longer HRI exposure.
Two of these are X\,3 and X\,4 of Johnston et al., identical to
the Einstein sources A and B detected near this cluster by Hertz \&\ Grindlay
(1983). X\,4/B is in the cluster core.
The HRI observations indicate that X\,3/A is extended,
as already found from the Einstein data (Hertz \&\ Grindlay 1983).
The positions of these sources as provided by the HRI are more
accurate than the PSPC positions, but the absence of an optical identification
leaves the overall uncertainty rather large, at 5$''$ (1-$\sigma$).

If X\,4/B is associated with the cluster its X-ray spectrum should be 
absorbed. 
Indeed, the PSPC countrate in channels 10-240 is 6.6$\pm$1.1\,\ctks, compatible
with the countrate given in 
Table\,\ref{tsour} for the hard band only (channels 50-240).
The HRI countrate is rather lower than one would expect on the basis
of the PSPC countrate, by about a factor three. The upper limit of the 
shorter HRI observation is compatible with the detected countrate in the
longer HRI observation, which indicates that the source dropped
in luminosity between the PSPC observation in March 1991 and the first
HRI observation of September 1992.

\subsection{NGC\,6715}

The cluster is in the field of view of a PSPC observation of the cataclysmic
variable V1223\,Sgr, but it is not detected. Due to the large offset, the
upper limit from the ROSAT All Sky Survey is but little improved upon.

\subsection{NGC\,6723}

The cluster has been observed twice with the HRI and is in the field of
view of several PSPC observations, of which I have analysed only the
longest one, which has the smallest offset. 
The cluster is detected in none of these images.
The HRI upper limit given in Table\,\ref{tsum} is based on the co-added
image.

\subsection{NGC\,6760}

The globular cluster is the target of one PSPC observation, and in the
field of view of two pointings at the transient X-ray source Aql X-1
(Johnston et al.\ 1996). The upper limit in Table\,\ref{tsum} is
from the co-added image of the two offset PSPC observations.

\subsection{NGC\,6779}

The cluster is not detected in the single HRI observation made of it.

\subsection{NGC\,6809/M\,55}

The PSPC observation has been published by Johnston et al.\ (1996); four
sources detected with the PSPC are also found with the longer HRI oservation.
One of these, X\,9, is within the (large) cluster core, and may belong to
the cluster. The HRI positions are more accurate than the previously published
positions based on the PSPC, but in the absence of an optical identification
the bore sight cannot be determined accurately.

\subsection{NGC\,6838/M\,71}

Two HRI observations have been made, pointed at the globular cluster.
Ten sources are detected, none within the half-mass radius of the cluster in 
the separate observations, nor in the added image.  A PSPC observation
pointed at 9\,Sge has NGC\,6838 in its field of view. No source is
detected at the position of the cluster.

\subsection{NGC\,7089/M\,2}

The cluster is not detected in the single HRI observation made of it.

\subsection{NGC\,7099/M\,30}

This cluster is the target of two PSPC observations, and in the field
of view of a much longer observation of the galaxy cluster MS\,2137.3$-$2353.
A source in the globular cluster is detected in all three observations, at 
countrates
consistent with a constant value. The numbers given in Table\,\ref{tsour} are
from the combined image of the shorter observations directed at the cluster.
One new source, X\,13 not related to the cluster, is detected with respect to 
the analysis by Johnston et al.\ (1994).
The cluster source has a countrate in channels 11-240 of 11.0$\pm$1.8\ctks,
higher than in the hard channels only (Table\,\ref{tsour}); this indicates 
a soft spectrum.
With $\nh=1.6\times10^{20}$\,\cmsq, as estimated from the visual reddening
of the cluster,
a bremsstrahlung spectrum of $kT\gtap0.8$\,keV gives the observed ratio 
of countrates.

\subsection{NGC\,7492}

The cluster is in the field of view of a PSPC observation of
Gliese\,890, but is not detected.

\section{Discussion}

The results of the analysis of the ROSAT data of globular clusters that do
not contain a bright X-ray source, i.e.\ that only contain dim sources 
if any, are summarized in Tables\,\ref{tsum} and \ref{tlist}.
I have added to these Tables the results for NGC\,5139 ($\omega$\,Cen),
NGC\,6397, NGC\,6752, and Liller\,1 from Johnston \&\ Verbunt (2000) and
for NGC\,6440 from Verbunt et al.\ (2000). \nocite{vkzh00}
The PSPC countrates of the central sources in NGC\,6397 and NGC\,6752 are
obtained from a reanalysis of the observations discussed in Johnston et al.\
(1994).
In this final section of the paper I compare the ROSAT results with
the Einstein observations, and then discuss the probability of 
spurious identifications, the spectra, and the luminosity 
distribution of the dim sources.

\subsection{Comparison with Einstein observations}

\begin{figure}
\centerline{\psfig{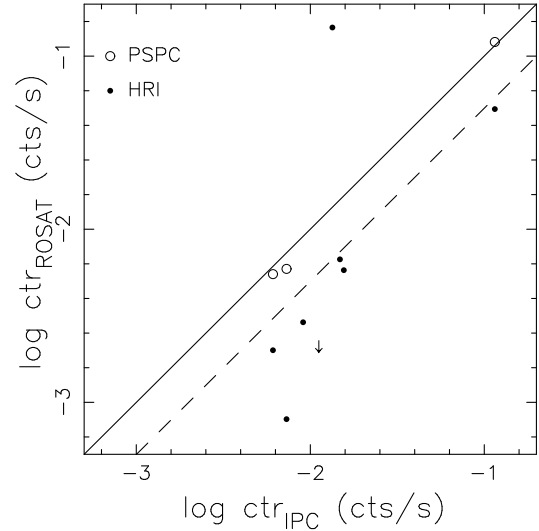}}
\caption{Comparison of the countrates found with ROSAT and with the
Einstein IPC, for PSPC observations (channels 50-140, $\circ$) and
for HRI observations ($\bullet$; upper limit shown as $\downarrow$).
The solid and dashed line shows the loci where the ROSAT countrates are 
equal or half the Einstein IPC countrates, as roughly expected for
PSPC and HRI observations, respectively.
\label{frosein}}
\end{figure}

All eight clusters near which a low-luminosity source was discovered
with the Einstein satellite have been observed with the ROSAT HRI;
three have been observed with the ROSAT PSPC.  The more accurate
positions obtained with ROSAT shows that all sources are within the
cluster core, with the exception of the source near NGC\,1904, which
is well outside the core.  Figure\,\ref{frosein} shows the comparison
between the Einstein IPC countrates (from Table\,4 of Hertz \&\
Grindlay 1983) and the ROSAT countrates as given in Table\,\ref{tsum}
(Table\,\ref{tsour} for NGC\,1904).  The three PSPC observations
all resulted in a detection, at a countrate (in channels 50-240)
similar to the Einstein countrate. This is remarkable, given that
at least two of these clusters show variable X-ray luminosities:
NGC\,104 (Verbunt \&\ Hasinger 1998) and NGC\,6656 (Sect.\,\ref{s6656} above).
Seven of the HRI observations detect the Einstein source. Of these
five detect the source at an HRI countrate roughly half that of the
Einstein IPC countrate, as expected for a constant source, or slighly 
below it. Two exceptions are the variable sources in NGC\,5272 
and NGC\,6656 (see Sects.\,\ref{s5272} and \ref{s6656} above).
The source in NGC\,5824 is not detected with the HRI, and apparently
was at a lower luminosity when ROSAT observed it in 1991 and 1994 than
during the Einstein observation.

\begin{table}[]
\caption[o]{Results of ROSAT observations of sources in the cores of globular 
clusters, and of sources within 10$''$ from the cluster center.
For each core the countrate in the PSPC
(channels 50-240) and HRI from pointed observations are given, as well 
as the X-ray luminosities in the 0.5-2.5\, keV band for the ROSAT All 
Sky Survey, PSPC pointings, and HRI pointings, for assumed 0.9\,keV 
bremsstrahlung spectrum. $n$ is the number of sources detected by the standard
analysis in the HRI image (indicated with '), or in the inner area of the 
PSPC. PSPC observations at an offset of more than 20$'$ are marked
separately. \label{tsum}}
\end{table}
\setcounter{table}{3}
\begin{table}
\begin{tabular}{lr@{ }rr@{ }r@{ }rr}
cluster   & PSPC     & HRI      & RASS     & PSPC     & HRI      & $n$ \\
 NGC 104   &   120.7  &    49.4  &    33.59 &    33.56 &    33.47 &  6' \\
 NGC 288   &          &     1.6  & $<$32.91 &          &    32.49 &  0' \\
 NGC 362   &          &     2.8  & $<$32.95 &          &    32.81 &  2' \\
 NGC 1261  & $<$ 2.3  & $<$ 0.9  & $<$33.48 & $<$32.94 & $<$32.72 &  2' \\
 Pal 2     &          & $<$ 0.8  & $<$34.80 &          & $<$34.06 &  1' \\
 NGC 1904  &          & $<$ 2.7  & $<$33.22 &          & $<$33.00 &  1' \\
 NGC 3201  & $<$ 2.8  &          & $<$32.52 & $<$32.19 &          &  3 \\
 NGC 4372  & $<$ 0.2  &          & $<$32.84 & $<$31.24 &          & 12 \\
 NGC 4590  &          & $<$ 1.0  & $<$33.28 &          & $<$32.52 &  0' \\
 NGC 5053  &          & $<$ 1.3  & $<$33.38 &          & $<$33.02 &  4' \\
 NGC 5139  &     5.5  &     2.0  & $<$32.58 &    32.42 &    32.36 & 12' \\
 NGC 5272  &          &   146.3  & $<$33.47 &          &    34.54 &  5' \\
 NGC 5286  &          & $<$ 1.4  & $<$33.82 &          & $<$32.96 &  2' \\
 NGC 5466  & $<$ 1.6  & $<$ 0.7  & $<$33.38 & $<$32.81 & $<$32.57 &  4' \\
 NGC 5824  &          & $<$ 2.1  & $<$34.49 &          & $<$33.97 &  0' \\
 NGC 5904  & $<$ 3.4  & $<$ 0.4  & $<$32.87 & $<$32.45 & $<$31.79 &  9' \\
 NGC 5986  &          & $<$ 3.1  & $<$33.23 &          & $<$33.29 &  0' \\
 NGC 6093  &     3.3  &          & $<$33.49 &    32.80 &          & 16 \\
 NGC 6121  & $<$ 2.1  & 0.5$^a$  & $<$32.13 & $<$31.41 &    31.20$^a$ &  4' \\
 NGC 6139  & $<$18.1  &     1.4  & $<$33.67 & $<$33.88 &    33.19 &  2' \\
 NGC 6205  &     2.4  &     2.0  & $<$32.83 &    32.31 &    32.47 & 12' \\
 NGC 6254  & $<$ 1.6  &          & $<$32.54 & $<$31.84 &          &  9 \\
 NGC 6266  &    22.1  &          & $<$33.48 &    33.48 &          &  9 \\
 NGC 6273  &          & $<$ 1.7  & $<$33.46 &          & $<$32.93 &  0' \\
 NGC 6304  &          & $<$ 2.2  & $<$33.34 &          & $<$32.82 &  1' \\
 NGC 6316  &          & $<$ 1.2  & $<$33.68 &          & $<$33.06 &  1' \\
 NGC 6341  &     6.5  &     1.4  & $<$32.69 &    32.81 &    32.38 &  7' \\
 NGC 6352  & $<$ 1.2  &          & $<$33.02 & $<$31.90 &          & 10 \\
 NGC 6366  &     2.5  &          & $<$32.40 &    32.10 &          &  8 \\
 HP 1      &          & $<$ 2.9  & $<$33.59 &          & $<$33.44 &  1' \\
 Liller 1  &          & $<$ 0.6  & $<$34.47 &          & $<$33.55 &  1' \\
 NGC 6380  &          & $<$ 1.4  & $<$33.91 &          & $<$33.44 &  3' \\
 NGC 6388  & $<$18.0  &          & $<$33.62 & $<$33.80 &          &  b \\
 NGC 6397  &    26.8  &     7.8  & $<$32.47 &    32.43 &    32.29 & 14' \\
 Djorg 1   &          & $<$ 2.9  & $<$33.79 &          & $<$33.79 &  1' \\
 NGC 6440  &          &     2.9  & $<$33.46 &          &    33.51 &  2' \\
 Terzan 6  &          & $<$ 1.6  &    36.06 &          & $<$33.72 &  1' \\
 NGC 6453  & $<$ 4.7  & $<$ 2.9  & $<$33.54 & $<$33.31 & $<$33.52 &  2' \\
 NGC 6496  & $<$ 0.7  &          & $<$33.80 & $<$32.23 &          &  5 \\
 NGC 6522  & $<$ 1.3  &          & $<$33.53 & $<$32.36 &          & 15 \\
 NGC 6528  & $<$ 1.0  &          & $<$33.73 & $<$32.43 &          & 15 \\
 NGC 6540  & $<$11.5  &          & $<$32.81 & $<$32.72 &          &  b \\
 NGC 6541  &          &     6.7  & $<$33.66 &          &    33.16 &  2' \\
 NGC 6544  & $<$ 2.2  &          & $<$32.66 & $<$31.77 &          &  1 \\
 NGC 6553  & $<$ 9.9  &          & $<$33.34 & $<$33.10 &          &  b \\
 Terzan\,11  & $<$ 1.3  &          & $<$33.35 & $<$32.56 &          &  b \\
 NGC 6626  &    19.5  &    11.8  & $<$33.32 &    33.25 &    33.44 &  9' \\
 NGC 6638  & $<$ 1.9  &          & $<$33.52 & $<$32.55 &          &  b \\
 NGC 6642  & $<$ 2.5  &          & $<$33.20 & $<$32.60 &          &  b \\
 NGC 6656  &     5.9  &     0.8  & $<$32.31 &    32.18 &    31.72 &  3' \\
 NGC 6715  & $<$10.0  &          & $<$34.30 & $<$34.13 &          &  b \\
 NGC 6723  & $<$ 2.6  & $<$ 1.0  & $<$33.07 & $<$32.49 & $<$32.40 &  3' \\
 NGC 6752  &    14.7  &     3.6  & $<$32.74 &    32.55 &    32.24 & 13' \\
 NGC 6760  & $<$ 0.6  &          & $<$33.32 & $<$32.14 &          &  b \\
 NGC 6779  &          & $<$ 0.5  & $<$33.31 &          & $<$32.41 &  1' \\
 NGC 6809  &     1.2  &     0.5  & $<$32.44 &    31.74 &    31.71 &  4' \\
 NGC 6838  & $<$ 2.5  & $<$ 0.3  & $<$32.37 & $<$31.92 & $<$31.41 & 10' \\
 NGC 7089  &          & $<$ 1.1  & $<$33.01 &          & $<$32.69 &  2' \\
 NGC 7099  &     8.0  &          & $<$32.94 &    32.88 &          &  4 \\
 NGC 7492  & $<$ 0.6  &          & $<$33.87 & $<$32.75 &          & 12 \\
\multicolumn{7}{l}{$^a$marginal detection, $^b$offset $>20'$}
\end{tabular}
\end{table}

\subsection{Probability of spurious identifications with clusters}
\label{secprob}

As noted in the introduction, the number of faint X-ray sources detectable
with ROSAT is large, and we must consider the possibility of chance coincidence
between a fore- or background X-ray source and a globular cluster.

Let us start by considering the source near Pal\,2. 
I approximate the area covered by the HRI detector with a circle with a radius
$r_{\rm d}$. 
As can be seen in Fig.\,\ref{fig6205} $15'\ltap r_{\rm d}\ltap20'$.
The probability $p$ that one serendipitous source in the HRI observation
is at a distance $r<R$ to the cluster center thus is 
$p=(R/r_{\rm d})^2$.
However, as noted by Rappaport et al.\ (1994), one must take into account 
the fact that many images are analyzed, in one of which the coincidence is 
found.
From Table\,\ref{tsum} it is seen that 147 sources are found in 34 HRI
observations, not counting the observations with zero detections.
(For each cluster only one observation is considered,
viz.\ the one with the highest sensitivity. 
The numbers $n$ or $n'$ given in Table\,\ref{tsum} are from
that observation, which may be a co-added observation.)
Taking $R=30''$ and $r_{\rm d}=20'$, I find that the probability for
one trial to detect a source at $r<R$ is $p=0.000625$; and the probability
of finding no such source after 147 trials is $(1-p)^{147}\simeq91$\%.
From this it is likely that all sources detected in the listed HRI observations
within 30$''$ of the cluster center are indeed cluster members; including 
the source near Pal\,2.

Two more sources detected in HRI observations and indicated in 
Table\,\ref{tsour} (see also Table\,\ref{tlist})
as possibly related to a globular cluster are those near
NGC\,1904 and NGC\,6626, located 71$''$ and $43''$ from the cluster centers,
respectively; the probabilities of finding no source within $r<71''$
and $r<43''$ after 147 trials are 60\%\ and 83\%.  

The inner area of the PSPC has a radius $r_{\rm d}=20'$, and from 
Table\,\ref{tsum} it is seen that 104 sources are found in 12 PSPC 
observations of clusters in which no source was found with the HRI.
(I count here only those PSPC observations pointed at a direction within 20$'$
from the globular cluster; and note that NGC\,6522 and NGC\,6528 are
in the same observation.)
Two sources detected in these PSPC observations and indicated in 
Table\,\ref{tsour} as possibly related to a globular cluster are those near
NGC\,6388 and NGC\,6352, located 44$''$ and 90$''$
from the cluster centers, respectively. The probabilities that no source
source is detected after 104 trials within these three distances are
87\%\ and 56\%, respectively.
None of the globular cluster sources observed with the PSPC at an offset
of more than 20$'$ is detected; these observations do serve to improve
on the upper limits from the the ROSAT All Sky Survey.

Because the probability in most individual observations of detecting
no source near the center by chance is rather close to unity, the overall
probabilities for detecting no source near a cluster center by chance
in the whole data set do not change much if we remove some clusters
from the sample, or e.g.\ switch NGC\,104 from the HRI set of observations
to the PSPC set. In this respect, the probabilities just given are fairly
stable.
For the same reason, if we argue that the number of trials is actually
less than 147 for the HRI, or less than 104 for the PSPC because some
sources certainly belong to the clusters, the probabilities just given
do not increase significantly.
A more important uncertainty is the effective size of the detectors,
considering that sources are more easily detected near the center
of the detectors, where the point spread function is sharper. For bright
sources this makes no difference, for faint sources it does.
With somewhat smaller effective radii, the probabilities given above all
increase, but not so dramatically as to change the conclusions.
Thus, for $r_{\rm d}=15'$ the probability of finding no source by chance at 
$r<30''$ from the cluster centers in any HRI observation is 85\%;
and the probability of finding no source by chance at $r<90''$ from the
cluster center in the PSPC observations is 39\%.

In addition to the probabilities discussed so far, it is interesting 
to consider the probability that all sources found within a cluster core
are actually related to the cluster; because the apparent size of the core 
is different for each cluster, this means that the probability of chance 
coincidence varies between clusters.
Writing the core radius in observation $i$ as $r_i$ and the number
of sources detected in that observation as $n_i$, we may write the
probability of getting no chance coincidence in observation $i$ as
$$ P_{0i} = \left[1-\left({r_i/r_d}\right)^2\right]^{n_i} $$
and the probability of getting no chance coincidence in a set of observations
as
$$ P (0) = \Pi_i P_{0i}. $$
The probability of getting a single chance coincidence in a set of obserations
may be written
$$ P (1) = \sum_j\left[(1-P_{0j})\Pi_{i\neq j} P_{0i}\right] =
           \sum_j\left( {1\over P_{0j} } -1 \right) P(0) $$
With these equations, taking $r_{\rm d}=20'$, and using the core radii from 
the June 22, 1999 revision of the catalogue described by
Harris (1996), I find that the probability of getting zero or one chance 
coincidence in the observations for which $n_i$ is given in Table\,\ref{tsum}
are 65\%\ and 30\%, respectively, for the 34 HRI observations, and
78\%\ and 20\%\ for the 13 PSPC observations. (Here the PSPC observation 
containing NGC\,6522 and NGC\,6528 is counted twice.)
Using the half-mass radii rather than the core radii, I find that the
probabilities of getting zero or one chance coincidence is 15\%\ and 32\%\
for the HRI observations and 35\%\ and 42\% for the PSPC observations.
\nocite{har96}
These numbers too are uncertain mainly because of the use of a fixed
effective detector radius, independent of source brightness.

\begin{table}[]
\caption[o]{Sources detected outside cores of globular
clusters with ROSAT, but possibly associated with them.
For each source, numbered according to Table\,\ref{tsour}, the apparent
distance to the cluster center and the 0.5-2.5\,keV luminosity,
assuming a 0.9\,keV bremsstrahlung spectrum, are listed. Data for NGC\,104,
NGC\,6397 and NGC\,6752 are from Verbunt \&\ Johnston (2000); their
luminosities have been recomputed from the countrates.
\label{tlist}}
\begin{tabular}{l@{ }rrrl@{ }rrr}
NGC   & X & d($''$) & $\log L_{\rm X}$ &NGC   & X & d($''$) & $\log L_{\rm X}$ 
\\
 104 &  4 &  79 & 31.68 & 6205 &  F &  75 & 32.21 \\
 104 &  6 &  53 & 31.78 & 6352 &  4 &  90 & 32.10 \\
 104 & 11 &  58 & 31.55 & 6388 &  4 &  44 & 33.77 \\
 104 & 13 &  53 & 31.38 & 6397 & 12 &  45 & 31.00 \\
Pal\,2 & 1 & 29 & 34.33 & 6626 &  7 &  43 & 32.15 \\
1904 &  1 &  71 & 33.18 & 6752 &  4 &  48 & 32.17 \\
5139 &  7 & 264 & 32.21 & 6752 &  6 &  92 & 31.68 \\
5139 & 21 & 170 & 31.76 & 7099 &  8 &   6 & 32.88 \\
5904 &  4 & 290 & 31.79 \\
\end{tabular}  
\end{table}

Notwithstanding the uncertainties in the statistical arguments given above,
it appears that most of the sources detected with the HRI
and PSPC in cluster cores are actually associated with a globular cluster.
On the basis of their distances to the cluster center, the sources listed
in the cores of NGC\,5139, NGC\,6809 and NGC\,6366 have non-negligible 
probabilities
of being fore- or background objects, as do most out-of-the-core sources
listed in Table\,\ref{tlist}. However, it is worth of note that the source
in Table\,\ref{tlist} with the largest distance to a cluster center, the
one near NGC\,5904, is considered to be a secure member of that cluster,
which indicates that sources outside the cluster core should not be discarded
too easily as possible members.
The statistical arguments above indicate that some of the sources 
listed in Tables\,\ref{tsum} and \ref{tlist} may not be 
related to the clusters, but that it is very unlikely that many of them
are chance projections. 
Optical identifications are required to settle the question of membership 
for the individual sources.

\subsection{Spectral energy distributions}

The five cluster spectra with enough PSPC counts to allow a spectral fit
are shown in Fig.\,\ref{fspec}. 
The most remarkable property of these spectra is that all peak at
low energies, indicating that they are fairly soft.
This is true also for the spectrum of NGC\,104, which is the sum of
the spectra of several sources.
(For example, an 0.4 to 2\,keV blackbody spectrum in NGC\,104 would show
a count distribution peaking at channel 100 to 130, respectively; the
observed spectrum peaks around channel 75.)
The X-ray luminosities of these sources, as well as their soft
spectra, are similar to the luminosities and spectra of soft X-ray
transients in quiescence (see e.g. Verbunt et al.\ 1994; Asai et al.\ 1998).
Spectra of cataclysmic variables, on the other hand, tend to be
harder, with temperatures for thermal bremsstrahlung fits of 2.4\,keV on
average, but with a large spread (e.g. Richman 1996, Van Teeseling et al.\ 
1996).
\nocite{vbj+94}\nocite{adh+98}\nocite{ric96}\nocite{tbv96}

\begin{figure}
\centerline{\psfig{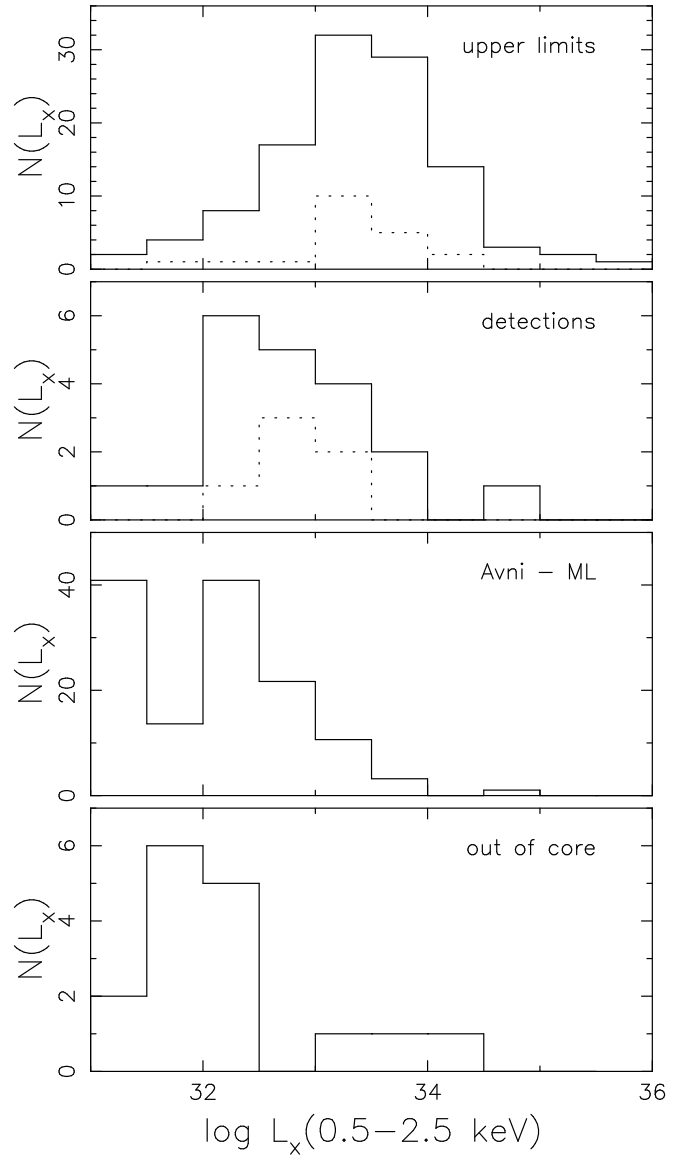}}
\caption{Top frames: Distributions of upper limits and detections of X-ray 
luminosities
of cores of globular clusters from ROSAT observations, derived assuming a
0.9\,keV bremsstrahlung spectrum. The solid lines are for all clusters, the
dotted line for collapsed clusters only.
The maximum-likelihood (Avni et al.\ 1980) luminosity
distribution derived from these observations is shown in the third frame.
The lowest frame gives the luminosity distribution for the sources
detected well outside the cluster cores.
\label{favni}}
\end{figure}
\nocite{astz80}

A problem with the fits for NGC\,104 and NGC\,6341 appears to be that they
underestimate the countrates at the lowest energies, even when for NGC\,6341
the absorption is set at a rather low value. The proximity of these sources
to the cluster centers (see Table\,\ref{tsour}) ensures that the sources
do belong to the cluster; and thus I conclude that the spectra of these sources
are softer than can be described by a single blackbody or bremsstrahlung 
spectrum.

Johnston et al.\ (1994) note that not all low-luminosity X-ray sources 
in globular clusters can have the same spectrum. The hardest
sources are those in NGC\,5139 ($\omega$\,Cen) and in NGC\,6752, but even for
these, Johnston et al.\ obtain low blackbody temperatures: 0.6$\pm$0.2 and 
0.35$\pm$0.05
keV, respectively. Note that in both clusters the PSPC spectrum is the
sum of spectra of at least two sources (Verbunt \&\ Johnston 2000).
Thus a range of spectral energy distributions is indeed observed, but
no hard spectra have been found amongst the dim sources so far.
As shown by Dotani et al.\ (1999) the highly variable source in NGC\,5272
is unique in changing its spectrum from extremely soft to fairly hard
as its luminosity increases (Sect.\ \ref{s5272}).

Observations with Chandra will be needed both to separate the spectra
of sources located close to one another in cluster cores, and to reduce the
problem of accurate background determination, which is difficult for 
possibly extended, faint sources.
The better spectral resolution will also allow comparison with more
realistic model spectra.

\subsection{Luminosity distribution}

\begin{figure*}
\centerline{
\parbox[b]{13.4cm}{\psfig{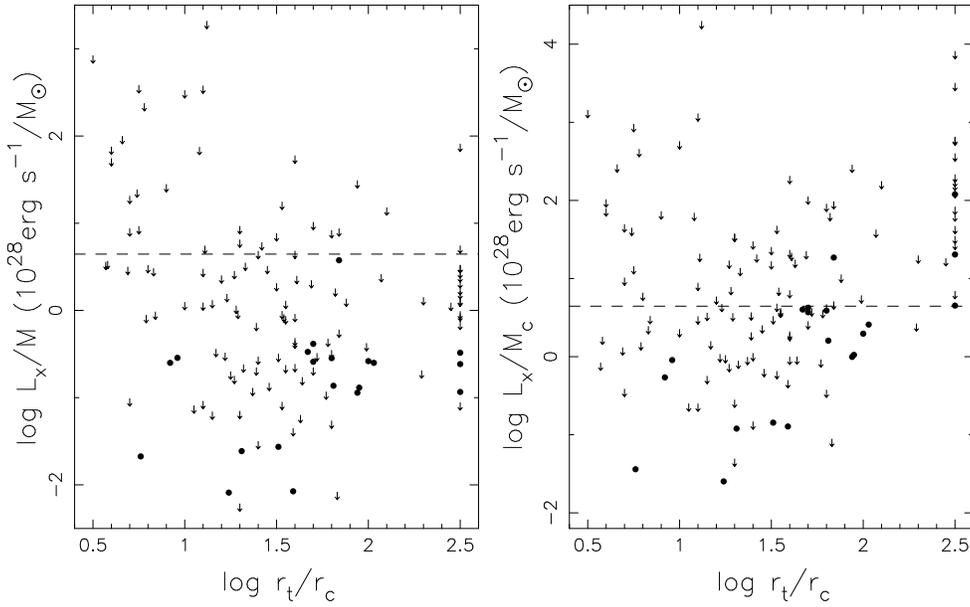}}
\parbox[b]{4.7cm}{\caption{0.5-2.5\,keV luminosity-to-mass ratio of globular clusters
as a function of the ratio of the tidal and core radii of the cluster. 
Collapsed clusters are located at $\log r_t/r_c=2.5$.
Left: the ratio for the total mass of the cluster,
right: the ratio for the core mass only.
To convert ROSAT countrate to 0.5-2.5\,keV luminosity an 0.9\,keV 
bremsstrahlung spectrum has been assumed. For comparison,
the horizontal dashed line gives the luminosity to mass ratio for the old 
galactic cluster M\,67.
\label{fxfopt}}}
}
\end{figure*}

For the study of the X-ray luminosity distribution of the dim sources
I start with Table\,\ref{tsum}. 
I retain for Terzan\,6 only the upper limit, i.e.\ 
ignore the detection of the transient in outburst.
For systems not listed in Table\,\ref{tsum} I use the upper limits derived
from the ROSAT All Sky Survey (Verbunt et al.\ 1995).
All these luminosities refer to the ROSAT range of 0.5-2.5\,keV, to
minimize the effect of the X-ray spectral energy distribution
on the conversion of the observed countrate to the X-ray luminosity.

For systems listed in Table\,\ref{tsum} I select the PSPC detection when
available; or else the HRI detection; or else the deepest from the upper
limits derived from the All Sky Survey or pointed PSPC and HRI
observations.
All count rates are converted to luminosities assuming an 0.9\,keV
bremsstrahlung spectrum, and using distances and absorption to the cluster from
Harris (1996; revision of June 22, 1999).
The resulting distributions of upper limits and detections are shown
in Fig.\,\ref{favni}, as is the maximum-likelihood intrinsic
luminosity distribution derived from these according to the method
described by Avni et al.\ (1980).
In this method the lowest bin (in this case for 
$\log L_{\rm x}{\rm (erg s}^{-1}{\rm )}<31.5$) contains all systems
in that and lower bins. From Fig.\,\ref{favni} we thus see that
about 60\%\ of all cluster cores have an X-ray luminosity above
$3\times10^{31}$\,\ergs.
This implies that the luminosity function cannot rise further towards 
lower luminosities, and on the contrary must drop rapidly below
$10^{31}$\,\ergs.

The distributions of upper limits and detections are indicated separately
in Fig.\,\ref{favni} for the cores which according to Harris (1996) are
collapsed or possibly collapsed. It is seen that these distributions
do not differ significantly from those of the total distribution.

I plot the integrated core luminosity as derived from detected sources,
because in crowded cores and/or in small cores ROSAT cannot separate
multiple sources. In the largest cores the luminosity given in 
Table\,\ref{tsum} and shown in Fig.\,\ref{favni} doesn't include the
sources below the ROSAT detection limit, and thus is a lower limit to
the actual core luminosity.
However, the argument given by Johnston \&\ Verbunt (1996)
that the total core luminosity is dominated by the brighter sources
still holds, and thus the estimates for the total luminosity are
expected to be correct within a factor 2.
\nocite{jv96}

Fig.\,\ref{favni} also shows the luminosity distribution of the sources
outside the cores; this distribution is shifted by a about a factor 3
with respect to the distribution of the core luminosities. As the sources
outside the cores are most likely single, the shift may be due simply
to the fact that the core luminosities are often those of several sources.

Fig.\,\ref{fxfopt} shows the ratio of X-ray luminosity to mass of the
globular cluster as a whole, and to its core mass, as a function of the 
ratio of the tidal radius $r_t$ to the core radius $r_c$ of the cluster.
The masses of the clusters and their cores are computed using the central 
luminosity densities and core radii tabulated by Harris (1996), and a 
mass-to-light ratio of 3.
For comparison the ratio of the old open cluster M\,67 is also shown.
For the X-ray luminosity of M\,67 I add the 0.1-2.4\,keV luminosities given 
in Table\,3 of Belloni et al.\ (1998), and multiply the sum by 0.4 to convert
to the 0.5-2.5\,keV bandpass; for the mass of M\,67 I use 723\,$\msun$
(Montgomery et al.\ 1993).\nocite{bvm98}\nocite{mmj93}
It should be noted that the core masses of globular clusters generally
are not accurately determined; mainly because low-mass stars can add
mass without adding luminosity.

Remarkably, most clusters or cluster cores do not emit more X-rays per
unit mass than the open cluster M\,67. 
The X-ray luminosity-to-mass ratio for the cluster as a whole is not
a function of its concentration, but there is a hint in Fig.\,\ref{fxfopt}
that concentrated clusters (with large $r_t/r_c$ values) emit more
X-rays per unit mass.  Whether this is because the sources in them are
more luminous or because they contain more sources cannot be
determined on the basis of the ROSAT data. It is true that two of the
three collapsed clusters (shown at $\log r_t/r_c=2.5$ in
Fig.\,\ref{fxfopt}), viz.\ NGC\,6397 and NGC\,6752, are known to
harbour at least four sources each in their cores (Verbunt \&\
Johnston 2000); whereas the PSPC observations obtained of the third
one, NGC\,7099, cannot resolve its small core.  On the other hand, in
cluster cores that are large enough to be completely resolved by
ROSAT, as is the case in $\omega$\,Cen, faint sources are not detected
individually, and for these clusters the total core luminosity is
underestimated.

\acknowledgements
I thank Lucien Kuiper for help in making Fig.\,\ref{fpuls}d.
This research is based on X-ray data from the ROSAT Data Archive of the
Max Planck Institute f\"ur extraterrestrische Physik at Garching, and
made extensive use of the SIMBAD database operated at Centre de Donn\'ees
astronomiques in Strasbourg. Indeed, without these data bases the research
described in this paper would have been impossible.

\appendix
\section{Expected X-ray flux for main-sequence stars, and field stars
detected in ROSAT observations of globular cluster.}

\begin{figure*}
\centerline{\psfig{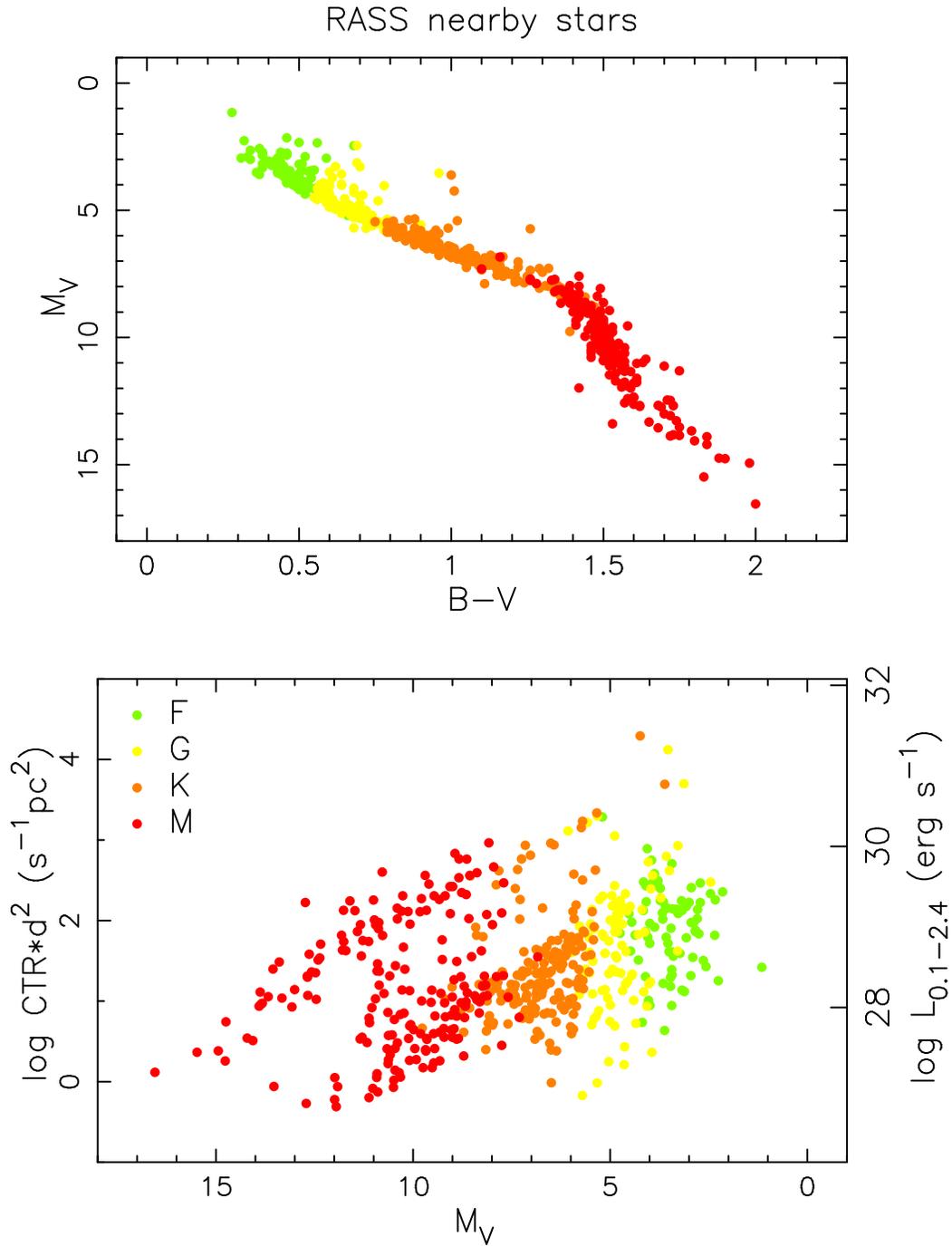}}
\caption{Above: location in the Hertzsprung Russell diagram of nearby
stars uniquely identified as counterparts of X-ray sources in the ROSAT All
Sky Survey; and not indicated as (sub)giant in their spectral type. 
Below: the product of ROSAT PSPC countrate and the distance
squared (as measure of the absolute X-ray luminosity in the ROSAT
band of 0.1-2.4\,keV) as a function of absolute visual magnitude, for
the same stars shown in the top graph. An approximate X-ray luminosity
is indicated on the right. Data for this Figure are from H\"unsch et al.\ 
(1999).
\label{frass}}
\end{figure*}

To get an impression of the X-ray fluxes that can be expected for
main-sequence stars, I use data from the ROSAT All Sky Survey for
nearby stars, as published by H\"unsch et al.\ (1999).  In many cases
(in particular for visual binaries)
the X-ray flux cannot be assigned uniquely to an optical counterpart,
and H\"unsch et al.\ (1999) list several possibilities.  From their
Table I select only unique identifications, which have a measured
value for colour $B-V$ and distance, and remove those with spectral
luminosity classes III or IV, or with absolute magnitude $M_{\rm V}<1$.  
This leaves mainly nearby main-sequence
stars.  From the observed visual magnitude $V$, the ROSAT PSPC countrate
CTR (\cts), and the distance $d$ (pc) I compute the absolute magnitude 
$M_{\rm V}$
and an absolute ROSAT countrate CTR\,$d^2$, neglecting interstellar
absorption.  The Hertzprung Russell diagram of the stars thus selected,
and a graph showing the absolute ROSAT countrate as a function of
absolute visual magnitude are shown in Figure\,\ref{frass}.  An
approximate X-ray luminosity scale is also indicated, based on a
conversion of 1\,\cts\ to $10^{-14}$\,\ergcms. 
(Note that H\"unsch et al.\ use a countrate to flux conversion factor 
which depends on the ratio of PSPC counts above and below 1\,keV, which 
implies that the X-ray luminosities derived by these authors may differ
somewhat from those according to Fig.\,\ref{frass}.) The Hertzprung Russell
diagram shows that only a few stars not on the main sequence passed
the selection procedure; and the lower graph that the range of X-ray
luminosity (in the ROSAT band) for each spectral type spans several
orders of magnitude.  Closer inspection of the Table in H\"unsch et al.\ 
(1999) shows that most of the stars
in the ridge that forms the higher envelope of the distribution, i.e.\
near the locus $\log({\rm CTR}d^2)= 0.25(17-M_{\rm V})$, are emission
line objects, as indicated by an 'e' in the spectral type.
Fig.\,\ref{frass} may be compared with Figures in which the ratio of
optical to X-ray flux for main-sequence stars is given (e.g.\
Hempelmann et al.\ 1995), but contains more information thanks to its 
use of the known distances. \nocite{hss+95}

\begin{table}[]
\caption[o]{Field stars detected in globular cluster observations.
Visual magnitude, colour and spectral type are from the Hipparcos
and Tycho Catalogues, or from SIMBAD. 
Colons mark spectral types derived from the observed
colours  with the assumption of zero reddening.
Absolute magnitudes are computed from Hipparcos paralaxes when available;
those indicated with a colon are derived from the spectral type.
The countrates (per second) are for channels 11-240 of the PSPC when no 
indication of interstellar absorption is found; otherwise twice the countrate
listed in Table\,\ref{tsour} from HRI or PSPC channels 50-240 is taken
for the computation of $M_{\rm X}\equiv \log {\rm CTR}d^2$.
\label{tstar}}
\begin{tabular}{l@{ }lr@{ }r@{ }l@{ }rr@{ }r}
NGC  & star & $V$\phantom{00} & $B$$-$$V$ & SpT & $M_{\rm V}$ & CTR & 
$M_{\rm X}$ \\
6093 & \multicolumn{6}{l}{VV\,Sco$^a$} \\
     & HD\,146457        &  8.46 & 0.32 & A5III & 0.3:$^b$ & 2*P & 3.1 \\
     & HD\,146516        & 10.14 & 0.82 & G1V   & 4.6:$^b$ & 2*P & 3.9 \\
6121 & HIP\,80290        &  9.72 & 0.63 & G0V   & 5.2      & 2*H & 2.9 \\
6266 & TY\,7360\,394 & 10.1\phantom{0} & 1.0\phantom{0} & K0V: & 5.9: & 0.235 & 3.1 \\
6341 & \multicolumn{6}{l}{V798\,Her$^c$} \\ 
     & TY\,3081\,510        & 11.0\phantom{0} & 0.69 & G5V: & 5.2: & 0.107 & 3.3\\
6352 & HIP\,85326        &  8.86 & 0.78 & K1V & 5.5 & 0.049 & 2.0 \\
     & HIP\,85342        &  7.05 & 0.44 & F6V & 3.7 & 0.056 & 2.1 \\
6366 & HIP\,85365        &  4.53 & 0.39 & F3V & 2.1 & 0.124 & 2.0 \\
6388 & TY\,7896\,3885 & 10.2\phantom{0} & 0.7\phantom{0} & G5V: & 5.2: & 
0.270 & 3.4 \\
     & HIP\,86356        &  8.12 & 0.42 & F5V & 2.2 & 0.120 & 3.4 \\
     & TY\,7896\,3812 & 10.5\phantom{0}& 0.9\phantom{0} & K1V:& 6.1: &
0.252 & 3.2 \\
     & TY\,7896\,2299 & 10.1\phantom{0}& 1.1\phantom{0} & $<$K5V &$<$7.3 &
2*P   & $>$1.7 \\
6541 & TY\,7911\,112 & 10.36 & 0.94 & K2V: & 6.3: & 2*H & 2.2 \\
6626 & TY\,6848\,3536 & 11.1\phantom{0}& 0.3\phantom{0}& F0 & 2.5: & 2*H & 2.5\\
\multicolumn{7}{l}{$^a$\,T\,Tau star; $^b$\,$A_{\rm V}\simeq0.2$;
$^c$\,W UMa star}\\
\end{tabular}  
\end{table}

For a suggested stellar counterpart to a ROSAT source in one of the images
discussed in this paper, I use spectral type and distance whenever known,
as derived from information available in SIMBAD.
When these quantities are not known, I derive an estimated spectral type
from the observed $B-V$ by assuming that reddening may be neglected, and
that the star is on the main sequence; and then estimate a distance
by comparing the absolute magnitude for a star of that spectral type
with the observed magnitude.
For this I use colours and absolute magnitudes of main sequence stars 
from Tables 3-2 and 3-3 of Mihalas \&\ Binney (1981).
\nocite{mb81}
The countrates used in this paper are HRI countrates or PSPC countrates
in channels 50-240, and must be converted to a total PSPC countrate
(channels 11-240) for comparison with the ROSAT All Sky Survey data.
I thus multiply the HRI or PSPC (ch.\ 50-240) countrates as listed 
in Table\,\ref{tsour} with a factor
2 to estimate the CTR value; for PSPC observations this value can be
checked directly from the analysis, provided that the absorption is
negligible, so that a sufficient number of counts is detected in channels
11-50. The factor 2 is approximate, but the uncertainty introduced is
negligible with respect to the range of X-ray luminosities at each
absolute visual magnitude.
Finally, I compare the values of CTR$d^2$ and $M_{\rm V}$ of the proposed
counterpart with Fig.\,\ref{frass}. In all cases the proposed counterpart
was found to have an acceptable X-ray luminosity in the ROSAT band.

Two stars are not main sequence stars, viz.\ the T\,Tau star VV\,Sco,
and the contact binary V798\,Her.
More information on the field stars can be found in the sections on
the individual clusters.

\end{document}